\documentclass[fleqn,usenatbib]{mnras}

\usepackage[T1]{fontenc}
\usepackage{ae,aecompl}

\usepackage{graphicx}	
\usepackage{amsmath}	
\usepackage{amssymb}	

\usepackage{xspace}
\usepackage{color}
\usepackage[normalem]{ulem}
\usepackage[dvipsnames]{xcolor}
\usepackage{soul}

\newcommand{\epic}{EPIC\,212803289\xspace}
\newcommand{\ktwo}{{\it K2}\xspace}
\newcommand{\kep}{\textit{Kepler}\xspace}
\newcommand{\feh}{\mbox{[Fe/H]}\xspace}
\newcommand{\teff}{\mbox{$T_{\rm \star, eff}$}\xspace}
\newcommand{\logg}{\mbox{$\log g_\star$}\xspace}
\newcommand{\vsini}{\mbox{$v \sin i_{\star}$}\xspace}
\newcommand{\kms}{\mbox{km\,s$^{-1}$}\xspace}
\newcommand{\ms}{\mbox{m\,s$^{-1}$}\xspace}
\newcommand{\mjup}{\mbox{$\mathrm{M_{\rm Jup}}$}\xspace}
\newcommand{\rjup}{\mbox{$\mathrm{R_{\rm Jup}}$}\xspace}
\newcommand{\mstar}{\mbox{$M_{\star}$}\xspace}
\newcommand{\rstar}{\mbox{$R_{\star}$}\xspace}
\newcommand{\densstar}{\mbox{$\rho_\star$}\xspace}
\newcommand{\msol}{\mbox{$\mathrm{M_\odot}$}\xspace}
\newcommand{\rsol}{\mbox{$\mathrm{R_\odot}$}\xspace}
\newcommand{\lstar}{\mbox{$L_{\star}$}\xspace}
\newcommand{\lsol}{\mbox{$\mathrm{L_\odot}$}\xspace}

\newcommand{\paperone}{Paper~I\xspace}
\newcommand{\tlcm}{{\sc tlcm}\xspace}
\newcommand{\target}{K2-99\xspace}
\newcommand{\planetb}{K2-99\,b\xspace}
\newcommand{\planetc}{K2-99\,c\xspace}

\usepackage{newtxtext,newtxmath}

\title[K2-99 revisited]{K2-99 revisited: a non-inflated warm Jupiter, and a temperate giant planet on a 522-d orbit around a subgiant}

\author[A. M. S. Smith et al.]
{A. M. S. Smith,$^{1}$\thanks{E-mail: Alexis.Smith@dlr.de},
S.~N.~Breton$^{2}$,
Sz.~Csizmadia$^{1}$,
F.~Dai$^{3}$,
D.~Gandolfi$^{4}$,
R.~A.~Garc\'ia$^{2}$ \newauthor 
A.~W.~Howard$^{5}$,
H.~Isaacson$^{6}$,
J.~Korth$^{7}$,
K.~W.~F.~Lam$^{8,1}$,
S.~Mathur$^{9,10}$,
G.~Nowak$^{9,10}$,\newauthor
F.~P\'erez Hern\'andez$^{9,10}$,
C.~M.~Persson$^{11}$,
S.~H.~Albrecht$^{12}$, 
O.~Barrag\'an$^{13}$,
J.~Cabrera$^{1}$,\newauthor
W.~D.~Cochran$^{14}$,
H.J.~Deeg$^{9,10}$,
M.~Fridlund$^{11,15}$,
I.~Y.~Georgieva$^{11}$,
E.~Goffo$^{4,16}$,\newauthor
E.~W.~Guenther$^{16}$,
A.~P.~Hatzes$^{16}$,
P.~Kabath$^{17}$,
J.~H.~Livingston$^{18}$,
R.~Luque$^{19}$, \newauthor
E.~Palle$^{9,10}$,
S.~Redfield$^{20}$,
F.~Rodler$^{21}$,
L.~M.~Serrano$^{4}$, and
V.~Van~Eylen$^{22}$
\\
$^{1}$Institute of Planetary Research, German Aerospace Center (DLR), Rutherfordstra{\ss}e 2, 12489 Berlin, Germany\\
$^{2}$AIM, CEA, CNRS, Universit\'e Paris-Saclay, Universit\'e Paris Diderot, Sorbonne Paris Cit\'e, F-91191 Gif-sur-Yvette, France\\
$^{3}$Division of Geological and Planetary Sciences, 1200 E California Blvd, Pasadena, CA, 91125, USA\\
$^{4}$Dipartimento di Fisica, Universit\'a di Torino, Via P. Giuria 1, I-10125, Torino, Italy\\
$^{5}$California Institute of Technology, 1200 E California Blvd., Pasadena, CA 91125, USA\\
$^{6}$University of California Berkeley, 501 Campbell Hall, Berkeley, CA 94709, USA\\
$^{7}$Department of Space, Earth and Environment, Astronomy and Plasma Physics, Chalmers University of Technology, 412 96 Gothenburg, Sweden\\
$^{8}$Center for Astronomy and Astrophysics, TU Berlin, Hardenbergstr. 36, 10623 Berlin, Germany\\
$^{9}$Instituto de Astrof\'\i sica de Canarias (IAC), 38205 La Laguna, Tenerife, Spain\\
$^{10}$Departamento de Astrof\'\i sica, Universidad de La Laguna (ULL), 38206 La Laguna, Tenerife, Spain\\
$^{11}$Department of Space, Earth and Environment, Onsala Space Observatory, Chalmers University of Technology, SE-439 92, Onsala, Sweden\\
$^{12}$Stellar Astrophysics Centre, Department of Physics and Astronomy, Aarhus University, Ny Munkegade 120, 8000 Aarhus C, Denmark\\
$^{13}$Sub-department of Astrophysics, Department of Physics, University of Oxford, Oxford, OX1 3RH, UK\\
$^{14}$Center for Planetary Systems Habitability and McDonald Observatory, The University of Texas at Austin,  Austin TX USA 78712.\\
$^{15}$Leiden Observatory, University of Leiden, PO Box 9513, 2300 RA, Leiden, The Netherlands\\
$^{16}$Th\"uringer Landessternwarte Tautenburg, Sternwarte 5, 07778 Tautenburg, Germany\\
$^{17}$Astronomical Institue, Czech Academy of Sciences, Fri\v{c}ova 298, 25165, Ond\v{r}ejov, Czechia\\
$^{18}$Department of Astronomy, University of Tokyo, 7-3-1 Hongo, Bunkyo-ky, Tokyo 113-0033, Japan\\
$^{19}$Instituto de Astrof\'isica de Andaluc\'ia (IAA-CSIC), Glorieta de la Astronom\'ia s/n, 18008 Granada, Spain \\
$^{20}$Astronomy Department and Van Vleck Observatory, Wesleyan University, Middletown, CT 06459, USA \\
$^{21}$European Southern Observatory, Alonso de C\'ordova 3107, Vitacura, Santiago, Chile\\
$^{22}$Mullard Space Science Laboratory, University College London, Holmbury St Mary, Dorking, Surrey RH5 6NT, UK
}

\date{Accepted 2021 November 26. Received 2021 November 26; in original form 2021 July 29}

\pubyear{2021}

\begin{document}
\label{firstpage}
\pagerange{\pageref{firstpage}--\pageref{lastpage}}
\maketitle

\begin{abstract}

We report new photometric and spectroscopic observations of the \target planetary system. Asteroseismic analysis of the short-cadence light curve from \ktwo's Campaign 17 allows us to refine the stellar properties. We find \target to be significantly smaller than previously thought, with \rstar = $2.55\pm0.02$~\rsol. The new light curve also contains four transits of \planetb, which we use to improve our knowledge of the planetary properties. We find the planet to be a non-inflated warm Jupiter, with $R_\mathrm{b} = 1.06 \pm 0.01$~\rjup. Sixty new radial velocity measurements from HARPS, HARPS-N, and HIRES enable the determination of the orbital parameters of \planetc, which were previously poorly constrained. We find that this outer planet has a minimum mass $M_\mathrm{c} \sin i_\mathrm{c} = 8.4\pm0.2$~\mjup, and an eccentric orbit ($e_\mathrm{c} = 0.210 \pm 0.009$) with a period of $522.2\pm1.4$~d. Upcoming TESS observations in 2022 have a good chance of detecting the transit of this planet, if the mutual inclination between the two planetary orbits is small.

\end{abstract}




\section{Introduction}

Giant planets found orbiting close to their stars have long been assumed to have formed beyond the snow line, and subsequently migrated towards their star, although \textit{in situ} formation has also been proposed (e.g. \citealt{Huang16,Batygin16}). Two classes of migration mechanism have been proposed to explain the existence of such planets: migration through the protoplanetary disc, and dynamical processes including planet -- planet scattering and Lidov-Kozai cycles, where eccentricity and inclination are exchanged periodically. Disc-driven migration \citep{G+T80, L+P86} is predicted to produce giant planets in circular orbits with low obliquities (i.e. the orbital axis and the axis of stellar rotation are well aligned). Planet -- planet scattering \citep{R+F96, W+M96} and migration via Lidov-Kozai cycles \citep{Lidov,Kozai,Eggleton01,W+M03,Fabrycky_Tremaine07}, however, should lead to highly eccentric orbits, with large obliquities. However, in the case of giant planets in orbits with periods of just a few days (the hot Jupiters), these orbital imprints of dynamical migration can be erased through tidal interactions with the host star. These tidal forces act to reduce the eccentricity and obliquity of the orbit, often on time-scales much shorter than the main-sequence lifetime of the host star.

Due to the strong dependence of these time-scales on the orbital distance ($\tau_e \propto a^{-13/2}$; e.g. \citealt{Jackson08}), planets orbiting just a little further out than hot Jupiters are thought to retain their primordial (post-migration) eccentricity and obliquity, because the time-scales for circularisation and alignment are longer than the stellar main-sequence lifetime. These warm Jupiters (usually defined as giant planets orbiting at distances greater than 0.1~AU, or with periods longer than 10~d) are intrinsically rarer than hot Jupiters \citep{Wittenmyer10, Santerne16}. They are also substantially more difficult to detect with ground-based, wide-field transit surveys such as WASP \citep{Pollacco06} or HAT \citep{HATnet}, which are responsible for the majority of hot Jupiter discoveries.

There is growing evidence (\citealt{Dawson_Johnson18} and references therein) that multiple migration mechanisms are required to explain the observed populations of hot and warm Jupiters. The existence of warm Jupiters on eccentric orbits, and with massive outer companions (such as the CoRoT-20 system; \citealt{corot20_deleuil, corot20_rey}) is evidence for high-eccentricity tidal migration, for instance. On the other hand, many warm Jupiters are found to have close planetary companions \citep{Huang16}, the presence of which is incompatible with high-eccentricity migration. Other systems, such as Kepler-419 \citep{kepler419_1, kepler419_2} appear to be ideal examples of post-Kozai migration, but have a mutual inclination between planetary orbits that is thought to be too small for Kozai migration to have taken place.

The study of systems containing a warm Jupiter is therefore vital for our understanding of planetary migration, and transiting systems are particularly valuable. Solving the full three-dimensional geometry of warm Jupiter systems with outer companions may prove crucial to understand the role of outer companions in high eccentricity migration.

The discovery of a planet orbiting \target (= \epic) was reported in \citet[hereafter `\paperone']{K299}. \target, a subgiant, was observed during \ktwo's Campaign 6, and found to host a massive ($M_\mathrm{b} = 0.97\pm0.09~\mjup$, $R_\mathrm{b} = 1.29\pm0.05~\rjup$) transiting planet in an eccentric ($e_\mathrm{b} = 0.19\pm0.04$) orbit, with a period of 18.25~d. A systemic radial acceleration of $-2.12\pm0.04$~$\mathrm{ms^{-1}d^{-1}}$ gave strong evidence for the presence of a third body in the system. In \paperone, we concluded that this third body was most likely to be a massive planet or brown dwarf orbiting with a period of several hundred days. If this third body has a high mutual inclination with \planetb, it could be responsible for the high-eccentricity migration of the inner planet to its current orbit.

The KESPRINT\footnote{\href{http://www.kesprint.science}{http://www.kesprint.science}} team has continued to monitor the radial velocity of \target, in order to determine the orbit of the outer body. In this paper, we present these radial velocity measurements, along with new photometric observations from \ktwo's Campaign 17. We describe these new observations in Section~\ref{sec:obs}, and perform a new characterisation of the properties of the star, using asteroseismology in Section~\ref{sec:stellar}. In Section~\ref{sec:tlcm} we perform joint modelling of the transit photometry and radial velocities to determine the parameters of both \planetb and \planetc. Section~\ref{sec:signals} describes a search for additional signals in the data, and we discuss the likelihood that the outer planet exhibits transits, and the prospects for observing such transits in Section~\ref{sec:transits_c}. Finally, we discuss the architecture of the \target system in Section~\ref{sec:architecture}, and present our conclusions in Section~\ref{sec:conclusions}.

\section{Observations}
\label{sec:obs}
\subsection{Spectroscopy}
\label{sec:obs:spectra}
\subsubsection{HARPS and HARPS-N}
We have continued to monitor the radial velocity (RV) of \target with spectroscopic observations using the HARPS and HARPS-N instruments. Seventeen new measurements\footnote{Conducted under programmes 097.C-0948(A), 099.C-0491(B), 099.C-0491(A), 0100.C-0808(A), 0101.C-0829(A), 60.A-9700(G), and 1102.C-0923(A).} were made between 2017 August 18 and 2019 May 23 (UT) with the HARPS spectrograph \citep[][$\lambda\, \in$\,(378--691)\,nm, $R\,\approx\,115\,000$]{Mayor2003}, mounted on the ESO 3.6-m telescope at La Silla Observatory, Chile. A further 24 measurements\footnote{Conducted under programmes A33TAC\_15, A34TAC\_10, OPT17A\_64, A35TAC\_26, OPT17B\_59, CAT17B\_99, CAT18A\_130, OPT18A\_44, A37TAC\_37, OPT18B\_52, and A38TAC\_26.} were made between 2017 April 1 and 2019 March 10 (UT) using HARPS-N \citep[$\lambda\, \in$\,(378--691)\,nm, R\,$\approx$\,115\,000]{2012SPIE.8446E..1VC}, mounted on the 3.58-m Telescopio Nazionale Galileo (TNG), at the Roque de los Muchachos Observatory on La Palma, Spain.

The exposure times varied from 1800 to 3600 seconds in the case of HARPS and from 1500 to 4000 seconds in the case of HARPS-N, depending on weather conditions and scheduling constraints, leading to a S/N per pixel of 25--74 at 550~nm and of 30--87 at 550~nm for HARPS and HARPS-N, respectively. HARPS spectra were extracted using the off-line version \texttt{HARPS\_3.8} of the DRS pipeline and HARPS-N spectra using off-line version \texttt{HARPN\_3.7} of the DRS \citep{2014SPIE.9147E..8CC}. In the case of both spectrographs Doppler measurements (absolute RVs) and cross-correlation function (CCF) activity indicators (FWHM and bisector spans) were measured by cross-correlating the extracted spectra with a G2 mask \citep{1996A&AS..119..373B}. Based on the prescription provided by \cite{2011arXiv1107.5325L} we also measured Mount-Wilson S-index ($\mathrm{S_{MW}}$) using our custom-developed code. In Table~\ref{tab:rvs1} we list all of the HARPS and HARPS-N RVs, including those previously reported in \paperone for the sake of completeness.

\subsubsection{HIRES}

The California Planet Search team took 19 spectra of \target between 2016 June and 2016 August, using the HIRES spectrograph at the W.~M.~Keck Observatory on Mauna Kea, Hawaii, USA. The RVs were analyzed with an iodine-free high S/N observation (~200 at 550~nm) taken with the B3 decker ($0.86^{\prime \prime} \times 14.0^{\prime \prime}$), allowing for removal of light from night sky emission lights and reflected moonlight. The observing procedures follow those described in \cite{Howard10}. A median exposure time of 426~s results in spectra with a median S/N of 103 per pixel and an internal uncertainty of 3.6~\ms. In Table~\ref{tab:rvs2} we list the 19  newly-obtained HIRES RVs, as well as those previously obtained from the Tull and FIES instruments, and reported in \paperone.

\subsection{\ktwo Campaign 17}
\label{sec:C17}
\target was originally observed by the \ktwo mission, the re-purposing of the \kep satellite to observe in the ecliptic plane \citep{K2}, in Campaign~6 (\paperone). \ktwo's Campaign~17 ran from 2018 March 02 to 2018 May 08, and overlaps significantly the Campaign~6 field, including \target. \target was observed as one of 179 short-cadence targets, meaning observations were conducted every minute instead of the usual 30~minutes for most targets. Four transits of \planetb were observed during this campaign.

We started from the pixel-level data downloaded from the Mikulski Archive for Space Telescopes (MAST) website. The loss of two reaction wheels degraded the pointing stability of the \kep spacecraft significantly \citep{K2}. The photometric measurements thus suffered from short-term (hours) systematic variations. To mitigate this systematic effect, we employed an approach similar to that of \citet{vburg}. Briefly, we put down a circular aperture of 4 pixel in radius around the brightest pixel in the image. We then computed the centre of light within the aperture after subtracting the background median. We then fitted a spline between the flux summed within the aperture and the position of the centre of light. The detrending was done by dividing the original flux with the best-fit spline variation. Our pipeline has previously been used to extract short-cadence light curve for other systems observed by \ktwo \citep{Dai_qatar2}.

\section{Stellar characterisation}
\label{sec:stellar}
\subsection{Spectral analysis}
We co-added all of the HARPS spectra (from \paperone and the newly-obtained spectra described in Sec.~\ref{sec:obs:spectra} and listed in Table~\ref{tab:rvs1}). This resulted in a spectrum with a signal-to-noise of around 228, higher than that used in \paperone. 

We  modelled the stellar effective 
temperature, \teff, the surface gravity, \logg, abundances, and line widths with 
\href{http://www.stsci.edu/~valenti/sme.html}{{\sc{sme}}} \citep[Spectroscopy Made Easy;][]{Valenti1996, pv2017} 
version 5.22, 
a spectral analysis  package that 
fits our co-added HARPS spectra   to    synthetic   spectra for a given set of parameters.
We used the Atlas12 \citep{Kurucz2013} atmosphere grids and extracted the required 
atomic and molecular line data from \href{http://vald.astro.uu.se}{VALD} \citep{Ryabchikova2015}. 
We used  spectral features sensitive to  photospheric parameters such as the broad line wings of 
H$_\alpha$ that was used to model  \teff, the line wings of 
the \ion{Ca}{I} 
$\lambda \lambda$6102, 6122, and 6162 triplet and the   \ion{Mg}{I}b  
$\lambda \lambda$5167, 5172, 5183 triplet
that were used to
model \logg. The abundances of Fe, Mg, and Ca relative to hydrogen, and 
the projected stellar rotational velocity, \vsini, were modelled 
from narrow lines between 6200 and 6600~\AA. 
We found [Ca/H]\,=\,$0.24\pm0.05$, [Mg/H]\,=\,$0.25\pm0.08$,  
[Fe/H]\,=\,$0.20\pm0.05$, and $V \sin i_\star = 9.8\pm1.0$~km~s$^{-1}$. 
We fixed the macro- and micro-turbulent velocities, 
 $V_\mathrm{mac}$ and $V_\mathrm{mic}$, to 5.8~\kms \citep{Doyle14} and 1.2~\kms \citep{bruntt10}, respectively. 
Our modelling suggests (by comparing \teff to the tabulation of \citealt{Pecaut_Mamajek}) that K2-99 is a  F9\,IV star, with an uncertainty smaller than one subclass. This is a slightly earlier spectral classification than the G0\,IV determined in \paperone.

\subsection{SED fit}

We used the publicly available software \href{https://github.com/jvines/astroARIADNE}{{\sc{ariadne}}} (Vines \& Jenkins, in prep.; \citealt{Acton_Mdwarf}; \citealt{ngts14}) to derive the stellar radius. In brief, {\sc ariadne} analyses the  spectral energy distribution (SED) by fitting  grids of stellar models to catalogue photometry, constrained by the {\it Gaia} parallax. The {\it Gaia} EDR3 $G$, $G_{\rm BP}$, and $G_{\rm RP}$, 2MASS $J$, $H$, and $K$, and {\it WISE} $W1$ and $W2$, and the  Johnson $B$ and $V$ magnitudes from APASS were fitted to  the {\tt {Phoenix~v2}} \citep{Husser2013}, {\tt {BtSettl}} \citep{Allard2012}, \citet{Castelli2003}, and \citet{Kurucz1993} atmospheric model grids. Priors on \teff, \logg, and \feh were taken from our {\sc sme}  model, and on the parallax from {\it Gaia} EDR3 ($\theta = 1.93\pm0.02$~mas). Reddening was accounted for, with $A_V$ limited to the maximum line-of-sight value from the SFD Galactic dust map \citep{Schlegel1998,Schlafly2011}.

The  resulting stellar radius, computed with Bayesian Model Averaging, is found to be  $2.64 \pm 0.06$~\rsol. Combining this result with the surface gravity, we find a mass of $1.46 \pm 0.15$~\msol. Note that the high precision of the parameters determined by {\sc ariadne} results from the Bayesian model averaging technique used to derive the uncertainties from the posterior parameter distribution. The distributions for each model are averaged, weighted by the relative probability of each model, leading to smaller uncertainties than those obtained from any single model \citep{Acton_Mdwarf}.

\subsection{Asteroseismology}
\label{sec:asteroseismology}
To perform an asteroseismic analysis, the K2 light curve was first optimised for this purpose. Large outliers were removed following \citet{2011MNRAS.414L...6G} and all the gaps were interpolated using a multi-scaled discrete cosine transform following inpainting techniques as described in  \citet{2014A&A...568A..10G} and \cite{Pires15}.

We analysed the power spectral density (PSD) of the asteroseismic optimised lightcurve in order to determine the global seismic parameters of the solar-like oscillations \citep[see e.g.][for more details]{2019LRSP...16....4G}. The first seismic parameter is the frequency of maximum oscillation power, $\nu_{\rm max}$, which has been shown to be related to the surface gravity of the star \citep{Brown91}. The second quantity that we can extract from the PSD is the mean large frequency separation, $\Delta \nu$, which is the distance in frequency between two modes of same degree and consecutive orders. This quantity is proportional to the square root of the mean density of the stars \citep{Kjeldsen_Bedding}. Two different methods were used to estimated $\Delta \nu$ and $\nu_{\rm max}$. We first applied the A2Z pipeline \citep{Mathur10}. The mean large frequency spacing is computed by taking the power spectrum of the power spectrum in boxes of 300\,$\mu$Hz allowing us to also compute thresholds to determine the confidence level of the detection \citep[see][for more details]{Mathur10}. The frequency of maximum power comes from the fit of a Gaussian function in the region of the modes after removing the fit of the convective background using two Harvey laws \citep{Harvey85}. We detected the modes with more than 95\% of confidence level. We obtained $\Delta \nu$= 40.35\,$\pm$\,0.84\,$\mu$Hz and $\nu_{\rm max}$= 660\,$\pm$\,16\,$\mu$Hz. 

The second method that was applied consists in fitting on the PSD a global p-mode pattern using the {\sc apollinaire}\footnote{\url{https://gitlab.com/sybreton/apollinaire}} MCMC peakbagging library \citep{Breton21}. The p-mode pattern equation is adapted from Eq. 27 from \citet{Lund17}:
\begin{equation}
    \nu_{n,\ell} = \left(n + \frac{\ell}{2}  + \epsilon \right) \Delta \nu - \delta \nu_{0\ell} - \beta_{0\ell} (n - n_\mathrm{max}) + \frac{\alpha}{2} (n - n_\mathrm{max})^2 \; ,
    \label{eq:tassoul_2nd}
\end{equation}
where $n$ and $\ell$ are the mode order and degree, respectively, $\epsilon$ is a phase shift and $\alpha$ the mode curvature. We define $\delta\nu_{0\ell}$ as \citep[see e.g.][]{Corsaro12}:
\begin{equation}
\begin{split}
    &\delta\nu_{00} = 0 \; , \\ 
    &\delta\nu_{01} = \frac{1}{2}(\nu_{n,1} - \nu_{n+1,0}) - \nu_{n,1} \; , \\ 
    &\delta\nu_{02} = \nu_{n,0} - \nu_{n,2}  \; ,
\end{split}
\end{equation}
$\alpha$ and $\beta_{0\ell}$ are the curvature terms on $\Delta\nu$ and $\delta\nu_{0\ell}$, respectively, while $n_\mathrm{max}$ is given by:
\begin{equation}
    n_\mathrm{max} = \frac{\nu_\mathrm{max}}  {\Delta\nu} - \epsilon \; .
\end{equation}

This second methodology provides $\Delta \nu$= 39.98 $\pm$ 0.40 $\mu$Hz and $\nu_{\rm max}$ = 673.13 $\pm$ 18.17 $\mu$Hz, in agreement with the values from the A2Z pipeline.

The MCMC process, implemented with the ensemble sampler of the {\sc emcee} library \citep{Foreman-Mackey13} is designed to sample the distribution of parameters $\theta$ of the posterior probability:
\begin{equation}
    p (\theta | \mathbf{S_x}) = \frac{p (\mathbf{S_x} | \theta) p (\theta)}{p (\mathbf{S_x})} \;,
\end{equation}
where $p (\theta)$ is the parameter prior distribution, $p (\mathbf{S_x})$ a normalisation factor. As the PSD follows a $\chi^2$ with two degrees of freedom \citep{Woodard84}, $p (\mathbf{S_x} | \theta)$ the likelihood function is given by:
\begin{equation}
    p (\mathbf{S_x} | \theta) = \prod\limits_{i=1}^{k} \frac{1}{S(\nu_i, \theta)} \exp \left[ - \frac{S_{x_i}}{S(\nu_i, \theta)} \right]  \; ,
\end{equation}
with $S$ and $S_x$ the ideal and observed spectrum, respectively. The chains are sampled with 500 walkers and 1000 steps. The first 50 steps have been removed in order to correctly take the burn-in phase into account. The final parameters are taken as the median of the sampled distribution (with the burn-in phase correctly taken into account) and their uncertainties as the largest value when considering the differences between the median and the 16th and 84th percentiles of the distribution, respectively.

\begin{figure}
 	\includegraphics[width=\columnwidth]{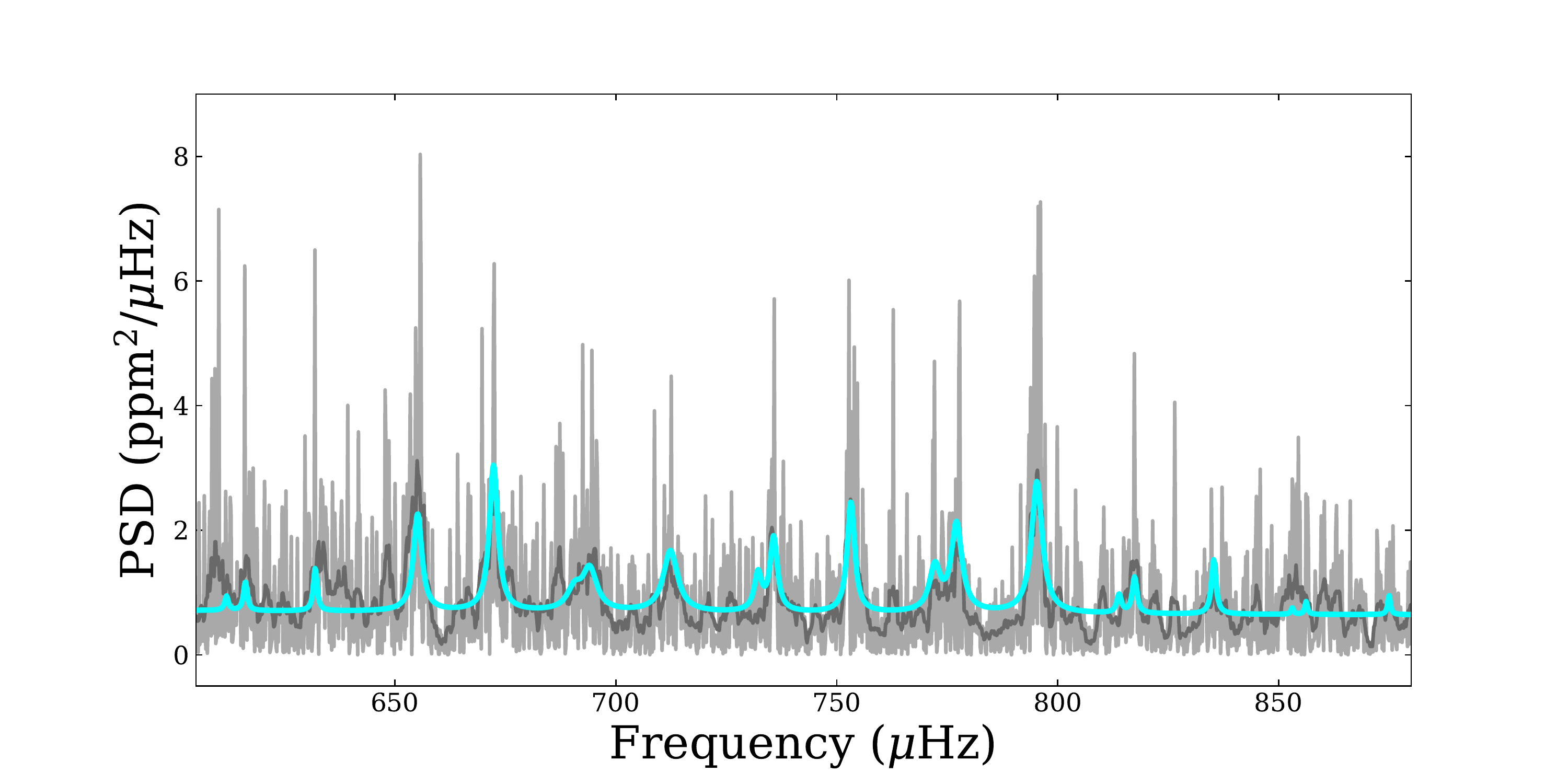}
     \caption{PSD of the asteroseismic optimised light curved (in light grey) at the natural resolution (no oversampling is performed) with a 11 points boxcar-smoothed version over-plotted in dark grey. The result of the {\sc apollinaire} fit is shown in cyan.}
     \label{fig:freqs}
 \end{figure}

To characterise the individual oscillation modes, {\sc apollinaire} fits a set of single Lorentzian profiles, one per each degree and radial order from $\ell=2$, $n=13$ up to $\ell=1$, $n=20$ (i.e. we did not try to fit either the inclination angle of the star, or for rotational splitting). Figure~\ref{fig:freqs} shows the result of the fit, overplotted on the PSD of the light curve optimised for asteroseismology. The frequencies of the 21 fitted modes are given in Table~\ref{tab:astero}.

Model fitting is based on a grid of stellar models evolved from the pre-main sequence to the RGB using the MESA
code \citep{Paxton11,Paxton13,Paxton15}, version 10\,398. 
The OPAL opacities \citep{Iglesias+Rogers96}, the GS98
metallicity mixture \citep{Grevesse+Sauval98} and the Eigenfrequencies were computed in the adiabatic approximation using the ADIPLS code \citep{ADIPLS}. The grid is composed of masses from $1.25\,M_{\odot}$ to $1.63M_{\odot}$ with a step of $\Delta M_{\odot} = 0.01M_{\odot}$, initial abundances [M/H] from $-0.10$ to $0.40$ with a step of $0.05$, mixing length parameters ($\alpha$) from $1.5$ to $2.2$ and step of $\Delta \alpha = 0.1$ and  overshooting parameter (for the Herwing prescription) $f_{ov}$ from $0$ to $0.04$ and step of $0.01$. The initial metallicity $Z$ and helium abundance $Y$ were derived from [M/H], constrained by taking a Galactic chemical evolution. Diffusion was not taken into account. As the 1D stellar evolution models do not properly model the outer turbulent layers of stars, we apply surface corrections.  More details can be found in \citealt{Perez-Hernandez19}.

\begin{figure}
 	\includegraphics[width=\columnwidth]{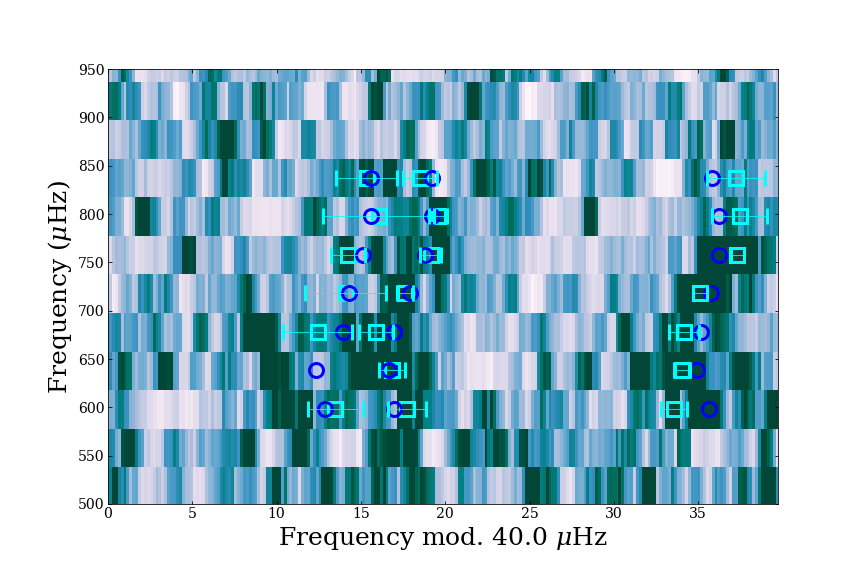}
     \caption{
       \'Echelle diagram, that is frequency as a function of the frequency modulo $\Delta \nu$ and with colours representing the power of the PSD, darker meaning higher power, of K2-99 in the region where the oscillation modes are detected. The cyan squares with symmetric error bars are the frequencies of the modes extracted. The blue circles represent the frequencies of the stellar
       model after a correction by surface effects as detailed in \protect\cite{Perez-Hernandez19} were applied. From left to right each vertical ridge corresponds to the modes $\ell$=2, 0, and 1.
     }
     \label{fig:ed}
 \end{figure}

A $\chi^2$ minimization, including p-mode frequencies and spectroscopic data, was applied to the grid of models. The procedure is described in \cite{Perez-Hernandez19}. The only difference is that here we have not used the luminosity derived from {\it Gaia} as an input parameter. To estimate the uncertainty in the output parameters we assumed normally distributed uncertainties for the observed frequencies, and for the spectroscopic parameters. We then search for the model with the minimum $\chi^2$ in every realization, and report mean and $1\sigma$ uncertainty values in Table~\ref{tab:stellar}. In addition we have done a $\chi^2$ minimization without considering the $\log g$ derived from the spectroscopic data but the results are the same within errors. Figure~\ref{fig:ed} shows the échelle diagram obtained by folding the PSD module $\Delta_\nu$. 
The stellar model p-mode frequencies are represented together with the fitted frequencies including the errors.

The results from our spectral analysis, SED fit and asteroseismic analysis are provided in Table~\ref{tab:stellar}, alongside the corresponding values from \paperone. The values from the various methods are in good agreement with each other (most are within 1 $\sigma$ of each other, and all are less than 2 $\sigma$ discrepant), and we adopt the asteroseismic stellar mass and radius values. Many of the stellar parameters computed here differ significantly from the \paperone values. Specifically, the star is significantly smaller, denser, and older than previously thought.

\begin{table*}
\caption{Stellar parameters for K2-99. As explained in Section~\ref{sec:asteroseismology}, we adopt the values of stellar mass and radius resulting from our asteroseismic analysis.}
\begin{tabular}{lc@{~~~~\vline~~~~}cccc}\hline
Parameter & unit & \paperone & {\sc sme} & SED & Asteroseismology\\
\hline
\teff & K & $5990 \pm 40$ & $6048 \pm 70$ & $6051 \pm 33$ & $6069 \pm 92$\\
\logg & [cgs] & $3.67 \pm 0.04$ & $3.77 \pm 0.10$ & $3.76 \pm 0.05$ & $3.783 \pm 0.004$\\
\vsini & \kms & $9.3 \pm 0.5$ & $9.8 \pm 1.0$ & -- & -- \\
\feh & dex & $+0.2 \pm 0.05$ & $+0.2 \pm 0.05$ & $+0.2 \pm 0.01$ & $+0.2 \pm 0.07$ \\
\rstar & \rsol & $3.1 \pm 0.1$ & -- & $2.64 \pm 0.06$ & $2.55 \pm 0.02$ \\
\mstar & \msol & $1.6 \pm 0.12$ & -- & $1.46 \pm 0.15$ & $1.44 \pm 0.03$ \\
\lstar & \lsol & -- & -- & $8.46 \pm 0.35$ & $7.93 \pm 0.46$ \\
 $A_V$ & mag & $0.05 \pm 0.05$ & $0.06 \pm 0.03$ & $0.03 \pm 0.01$ & --\\
Age & Gyr & $2.4 \substack{+0.2 \\ -0.6}$ & -- & -- & $3.7 \pm 0.4$ \\
\hline
\\
\end{tabular}
\label{tab:stellar}
\end{table*}

\subsection{{Stellar distance}}
The advent of {\it Gaia} EDR3 \citep{Gaia,GaiaEDR3}  and its exquisite parallax measurements allows us to improve upon the stellar characterisation of \paperone. \target can be found in the {\it Gaia} EDR3 catalogue with the identifier 3620612011248988416. A simple inversion of the parallax gives the distance to \target as $518\pm5$~pc. This is smaller than the value determined in \paperone ($606\pm32$~pc), but is consistent within $3\sigma$.

\section{Joint modelling of transit light curve and RVs}
\label{sec:tlcm}
\subsection{Method}
We combine the newly-obtained RVs with the RVs reported in \paperone  (Tables~\ref{tab:rvs1}~\&~\ref{tab:rvs2}), the C6 long-cadence {\sc k2sc} \citep{Aigrain16} light curve, and the new short-cadence C17 light curve (Section~\ref{sec:C17}) and fit them simultaneously. We perform the fit using the Transit Light Curve Modeller (\tlcm), which was previously used in \paperone (as well as in numerous other papers), and is now fully described by \cite{tlcm}.  In addition to the transit and radial velocity curve which we fitted for the inner planet in \paperone, we also fit a second Keplerian radial velocity component.

There is some residual correlated noise in the C17 light curve, evidenced by a slight variation in depth between transits, which may be the result of instrumental systematics, and/or stellar activity. We model this using the wavelet approach of \cite{wavelets}, as implemented in \tlcm \citep{tlcm, tlcm_red}. We are confident that this approach yields the correct transit depth, as our results are consistent with those obtained from fitting the C6 light curve alone, and with fitting the long-cadence C17 light curve produced with the {\sc everest} pipeline \citep{everest}.

We fit for a total of 22 parameters with \tlcm. For \planetb, we fit the orbital period ($P_\mathrm{b}$), the epoch of mid-transit ($t_\mathrm{0,b}$), the scaled semi-major axis ($a_\mathrm{b} / \rstar$), the ratio of planetary to stellar radii ($R_\mathrm{b} / \rstar$), the transit impact parameter ($b_\mathrm{b}$), two parameters relating to the orbital eccentricity $e_\mathrm{b}$ and argument of periastron $w_\mathrm{b}$ ($\sqrt{e_\mathrm{b}} \sin \omega_\mathrm{b}$ and $\sqrt{e_\mathrm{b}} \cos \omega_\mathrm{b}$), and the radial velocity semi-amplitude ($K_\mathrm{b}$). We also fit for the white noise ($\sigma_\mathrm{w}$) and red noise ($\sigma_\mathrm{r}$) levels in the light curve (defined as per \citealt{wavelets}), and the stellar limb-darkening parameters $u_+$ and $u_-$ (which are related to the quadratic coefficients $u_a$ and $u_b$ by $u_+ = u_a + u_b$ and $u_- = u_a - u_b$). For \planetc, we fit $P_\mathrm{c}$, the epoch of periastron ($t_{\rm peri,c}$), $\sqrt{e_\mathrm{c}} \sin \omega_\mathrm{c}$, $\sqrt{e_\mathrm{c}} \cos \omega_\mathrm{c}$, and $K_\mathrm{c}$. We also fit for the systemic radial velocity ($\gamma$), and four instrumental RV offsets ($\gamma_{\rm 2-1}$, $\gamma_{\rm 3-1}$, $\gamma_{\rm 4-1}$, $\gamma_{\rm 5-1}$).

\tlcm uses a Markov-chain Monte Carlo (MCMC) algorithm to sample the posterior parameter space. A total of 40 MCMC chains, each of 340\,000 steps were used, with the first 6\,000 steps discarded as burn-in. We used widely-spaced uniform priors on the fitted parameters, centred on the results from \paperone. Since the stellar density, $\densstar = 126\pm3.8$~kg m$^{-3}$, is determined from asteroseismology (Section~\ref{sec:asteroseismology}), we place a Gaussian prior on this quantity in the \tlcm fit.

\subsection{Results}
\begin{figure}
 	\includegraphics[width=\columnwidth]{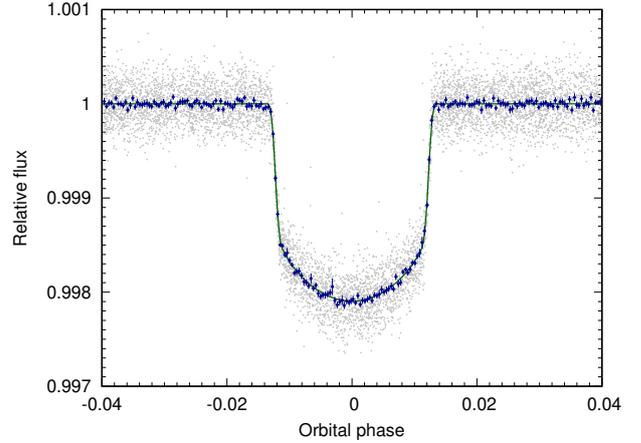}
     \caption{Phase-folded transit light curve. The full light curve from \ktwo Campaigns 6 and 17 is shown as grey circles, the same data binned in phase to the equivalent of 10 minutes is shown as blue squares. The wavelet model of the correlated noise is subtracted from the data, and our best-fitting model is indicated with a solid green line. }
     \label{fig:transit}
 \end{figure}

\begin{figure}
 	\includegraphics[width=\columnwidth]{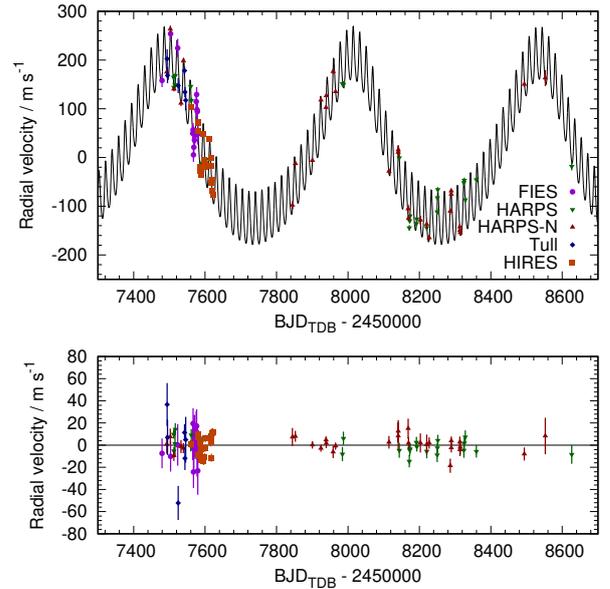}
     \caption{Radial velocity as a function of time. All 93 RV measurements (including those already presented in \paperone) are shown here. Measurements from FIES are represented by purple circles, HARPS by downward-pointing green triangles, HARPS-N by upwards-pointing red triangles, Tull by blue diamonds, and HIRES by orange squares. Our best-fitting model is shown as a solid black line. The residuals to this fit are shown in the lower panel.}
     \label{fig:rv_time}
 \end{figure}

\begin{figure}
 	\includegraphics[width=\columnwidth]{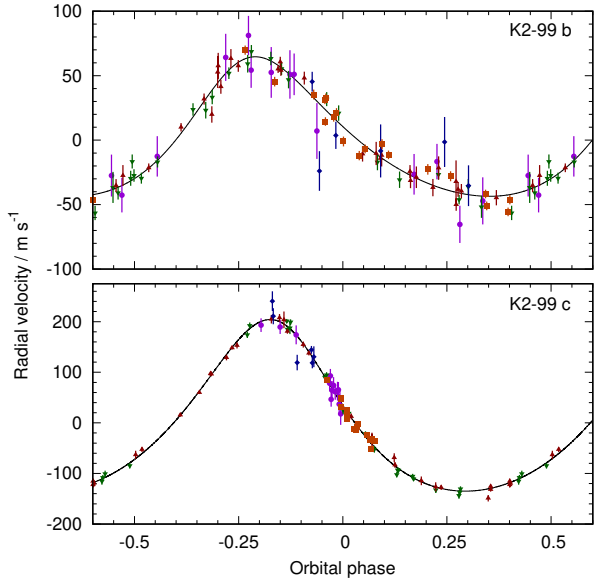}
     \caption{Radial velocity as a function of orbital phase for \planetb (upper panel) and \planetc (lower panel). In each panel, our best-fitting model is shown as a solid black line, and the colour and shape coding of the points is identical to Fig.~\ref{fig:rv_time}. Note that the orbital phase for planet `c' is computed with respect to the predicted time of mid transit for that planet.}
     \label{fig:rv_phase}
 \end{figure}

The resulting best fit (defined as the median of the MCMC posterior distribution) to the transit light curve of \planetb is shown in Fig.~\ref{fig:transit}. The radial velocity data are shown as a function of time in Fig.~\ref{fig:rv_time}, alongside the best-fitting model. Fig.~\ref{fig:rv_phase} shows the radial velocity data as a function of orbital phase for each planet, with the best-fitting model for the other planet subtracted from both the data and model in each case.

\subsubsection{Stellar density}

Fitting the data as described in the previous section, but omitting the prior on \densstar results in a best-fitting stellar density, $\densstar = 163 \pm 9$~kg m$^{-3}$, which is 3.8 sigma away from the asteroseismic value. The fitted parameters in the two cases are consistent within 1 sigma, with the exception of $b_\mathrm{b}$ and $a_\mathrm{b}/\rstar$, which differ by around 2 and 3 sigma, respectively.

\label{sec:results}
\subsubsection{\planetc}
We find that the outer planet has an orbital period $P_c = 523.1 \pm 1.4$~d, and an RV semi-amplitude, $K_c = 166\pm 2$~\ms, corresponding to a minimum mass, $M_c\sin i_\mathrm{c}$ of $8.2 \pm 0.2$~\mjup. This minimum mass is compatible with both a high-mass planet (for $i_\mathrm{c} \gtrsim 40\degr$), and a brown dwarf (for smaller values of $i_\mathrm{c}$). Like that of the inner planet, the outer planet's orbit is significantly eccentric, with $e_c = 0.211 \pm 0.009$. The orbital distance of $1.43\pm0.01$~au means that \planetc is too hot to be in the habitable zone, even according to the `optimistic habitable zone' of \cite{Kopparapu13}. This remains true, even if one considers the apastron distance of around 1.7~au, instead of the semi-major axis.

\subsubsection{\planetb}
\label{sec:results:planetb}
We improve our knowledge of the parameters describing \planetb, and present them alongside those derived in \paperone in Table~\ref{tab:mcmc}. The precision of the orbital period measurement is improved by a factor of around 30, allowing the transit time to be predicted to a precision less than 20 minutes for over two decades.

Most other parameters are in good agreement with the values from \paperone, but there are some notable exceptions. Our newly-determined value of the scaled orbital semi-major axis for the inner planet, $a_\mathrm{b} / \rstar$, is significantly larger than our previous measurement. The new value of $\omega_\mathrm{b}$ also differs by more than 2 $\sigma$ from the \paperone value. We re-examined our \paperone analysis, and conclude that these parameters were affected by a problem with our analysis in \paperone. We found that in our earlier analysis, the $e_\mathrm{b}\sin\omega_\mathrm{b}$ parameter\footnote{The version of \tlcm used in \paperone fitted the orbital eccentricity, $e_\mathrm{b}$ and argument of periastron, $\omega_\mathrm{b}$ through $e_\mathrm{b}\sin{\omega_\mathrm{b}}$ and $e_\mathrm{b}\cos{\omega_\mathrm{b}}$. However, in order not to inadvertently impose a non-uniform prior on $e$ (e.g. \citealt{Ford06}; \citealt{wasp30}; \citealt{Exofast}), the current version of \tlcm fits instead for $\sqrt{e_\mathrm{b}}\sin{\omega_\mathrm{b}}$ and $\sqrt{e_\mathrm{b}}\cos{\omega_\mathrm{b}}$.} did not converge properly. The current version of \tlcm uses the Gelman-Rubin statistic and estimated sample size to ensure that all parameters are well sampled, but these tests were not used in \paperone. This problem with $e_\mathrm{b}\sin\omega_\mathrm{b}$ resulted in a biased determination of $\omega_\mathrm{b}$. The eccentricity was not badly affected, since $e_\mathrm{b}\cos\omega_\mathrm{b}$ was determined correctly, and $e_\mathrm{b}\cos\omega_\mathrm{b} >> e_\mathrm{b}\sin\omega_\mathrm{b}$ in this case. However, in order to compensate for this incorrect $e_\mathrm{b}\sin\omega_\mathrm{b}$, the values of $a_\mathrm{b} / \rstar$ and $b_\mathrm{b}$ were also biased.

We also performed fits to just the C6 light curve, and just the C17 light curve and found in each case values of $a_\mathrm{b} / \rstar$ consistent with our new value in Table~\ref{tab:mcmc}. As an additional check, we also fitted the RVs with {\sc rvlin} \citep{rvlin}, and recovered values of $e_\mathrm{b}$, $\omega_\mathrm{b}$, $K_\mathrm{b}$, $P_\mathrm{c}$, $t_{\rm peri,c}$, $e_\mathrm{c}$, $\omega_\mathrm{c}$, and $K_\mathrm{c}$ in good agreement with our values from \tlcm.

Additionally, some uncertainties were underestimated in \paperone. Error estimation in the current version of \tlcm is performed by applying the `16-84 per cent' rule to the MCMC posterior distribution (as recommended by e.g. \citealt{Hogg+F-M}). Previously, the `varying $\chi^2$' method was used, which can lead to underestimated error bars, because it assumes that the uncertainties are Gaussian, and that the model is linear \citep{Andrae10}.

We also find a significantly smaller planet radius than previously; this is driven by the smaller stellar radius resulting from our new analysis (Section~\ref{sec:stellar}). The new planetary radius of $1.06\pm0.01$~\rjup indicates an uninflated planet, as expected from empirical studies of gas giants. \cite{Sestovic18}, for instance, found no evidence for an inflated population of planets in the mass range 0.37 -- 0.98 \mjup above an insolation of $1.6\times 10^6~\mathrm{W m^{-2}}$. \planetb receives less than half of this insolation, even when at periastron.

Finally, the stellar limb-darkening parameters, $u_+$ and $u_-$, also vary from those reported in \paperone. We tried fixing $u_+$ and $u_-$ to the values obtained in \paperone, and note no resulting difference to less than one $\sigma$ in any other parameters.

\begin{table*}
\caption{System parameters from \tlcm modelling} 
\label{tab:mcmc}
\begin{tabular}{lcccc}
\hline
\hline
Parameter                                   & Symbol                    & Unit                  & \cite{K299}           & This work  \\
\hline 
\planetb:                                   &                           &                       &                       &                                   \\
&\\
Orbital period                              & $P_\mathrm{b}$            & d                     & $18.249\pm0.001$      & $18.24783 \pm 0.00003$\\
Epoch of mid-transit	    	    	    & $t_{\rm 0,b}$             & $\mathrm{BJD_{TDB}}$  & $2457233.823\pm0.003$ & $2458182.7133 \pm 0.0005$\\
Transit duration                            & $t_{\rm 14,b}$            & d                     & $0.50\pm0.01$         & $0.462\pm0.007$\\
Scaled orbital semi-major axis              & $a_\mathrm{b}/\rstar$  & ...                   & {}$11.1\pm0.1${}      & $13.1\pm0.1$\\
Ratio of planetary to stellar radii         & $R_\mathrm{b}$ / \rstar   & ...                   & $0.0422\pm0.0006$     & $0.0426 \pm 0.0004 $\\
Orbital semi-major axis     	    	    & $a_\mathrm{b}$            & AU                    & $ 0.159\pm0.006$      & $0.153\pm0.001$\\
Transit impact parameter                    & $b_\mathrm{b}$            & ...                   & $0.41\pm0.05$         & $0.34\pm0.05$\\
Orbital inclination angle   	    	    & $i_\mathrm{b}$            & $^\circ$              & $ 87.7\pm0.3$         & $88.6\pm0.2$\\
...	                    & $\sqrt{e_\mathrm{b}} \sin \omega_\mathrm{b}$  & ...                   & ...                   & $ -0.20\pm0.03 $\\
...						& $\sqrt{e_\mathrm{b}} \cos \omega_\mathrm{b}$  & ...                   & ...                   & $ 0.42\pm0.01 $\\
Orbital eccentricity	    	    	    & $e_\mathrm{b}$            & ...                   & $0.19\pm0.04$         & $ 0.22\pm 0.01$\\
Argument of periastron                      & $\omega_\mathrm{b}$       &  $^\circ$             & $8\pm8$               & $334\pm4$\\
Stellar orbital velocity semi-amplitude     & $K_\mathrm{b}$            & m s$^{-1}$            & $56 \pm4 $            & $54 \pm 1$\\
Planet mass 	    	    	    	    & $M_\mathrm{b}$            & \mjup                 & $ 0.97\pm0.09  $      & $0.87\pm0.02$\\
Planet radius	    	    	    	    & $R_\mathrm{b}$            & \rjup                 & $1.29 \pm0.05  $      & $1.06 \pm 0.01$\\
log (planet surface gravity)     	        & $\log g_{\rm b}$          & (cgs)                 & $ 3.2\pm0.1  $        & $3.29 \pm 0.02$\\
Planetary equilibrium temperature$^\dagger$ & $T_{\mathrm{b},A=0}$      & K                     & ...                   & $1184 \pm 19$ \\
\hline
\planetc:                                   &                           &                       &                       &                                   \\
&\\
Orbital period	    	    	    	    & $P_\mathrm{c}$            & d                     & $485\pm310$           & $522.2 \pm 1.4$\\
Epoch of periastron	    	    	        & $t_{\rm peri,c}$          & $\mathrm{BJD_{TDB}}$  &...                    & $2458025.6\pm2.8$\\
Epoch of mid-transit$^\ddagger$	    	    & $t_{\rm 0,c}$             & $\mathrm{BJD_{TDB}}$  &...                    & $2458104.2 \pm 1.5$\\
Transit duration$^\ddagger$                 & $t_{\rm 14,c}$            & d                     &...                    & $<1.40 (3\sigma)$\\
Orbital semi-major axis     	    	    & $a_\mathrm{c}$            & AU                    & $1.4\pm1.0$           & $1.43\pm0.01$\\
...					    & $\sqrt{e_\mathrm{c}} \sin \omega_\mathrm{c}$  & ...                   & ...                   & $0.11\pm0.02$\\
...						& $\sqrt{e_\mathrm{c}} \cos \omega_\mathrm{c}$  & ...                   & ...                   & $0.45\pm0.01$\\
Orbital eccentricity	    	    	    & $e_\mathrm{c}$            & ...                   &                       & $0.210 \pm 0.009$ \\
Argument of periastron                      & $\omega_\mathrm{c}$       & $^\circ$              &                       & $13\pm2$ \\
Stellar orbital velocity semi-amplitude     & $K_\mathrm{c}$            & m s$^{-1}$            & $230\pm150$           & $170\pm2$\\
Minimum planet mass 	    	    & $M_\mathrm{c} \sin i_\mathrm{c}$  & \mjup                 & $14\pm9$              & $8.4\pm0.2$\\
Planetary equilibrium temperature$^\dagger$ & $T_{\mathrm{c},A=0}$      & K                     & ...                   & $390\pm 6$\\
\hline
Stellar parameters:                         &                           &                       &                       &                                   \\
&\\
Limb-darkening parameters                   & $u_+$                     & ...                   & $ 0.6\pm0.1$          & $0.41\pm0.05$\\
						                    & $u_-$                     & ...                   & $ 0.08\pm0.20$        & $0.7\pm0.2$\\
Photometric white noise & $\sigma_\mathrm{w}$ &...&...& $(174\pm1)\times 10^{-6} $ \\
Photometric red noise & $\sigma_\mathrm{r}$ &...&...& $(56\pm2)\times 10^{-4} $ \\
Systemic radial velocity     	    	    & $\gamma$                  & km s$^{-1}$           & $-2.08 \pm 0.01$      & $-2.855\pm0.005$\\
Velocity offset between FIES and HARPS      & $\gamma_{\rm 2-1}$        & m s$^{-1}$            & $  100\pm 8$          & $88\pm5$\\
Velocity offset between FIES and HARPS-N    & $\gamma_{\rm 3-1}$        & m s$^{-1}$            & $  110\pm 7$          & $98\pm5$\\
Velocity offset between FIES and Tull       & $\gamma_{\rm 4-1}$        & m s$^{-1}$            & $  3165\pm 12$        & $3154\pm9$\\
Velocity offset between FIES and HIRES      & $\gamma_{\rm 5-1}$        & m s$^{-1}$            &  ...                  & $2856\pm5$\\
\noalign{\smallskip}  
\hline
\end{tabular} \\ 
$^\dagger$ Assuming a planetary albedo of zero, and isotropic heat redistribution.
$^\ddagger$ If the outer planet does indeed transit.
\end{table*} 

\section{Search for additional signals in the data}
\label{sec:signals}
\subsection{Transit timing variations (TTV)}

\subsubsection{Predicted TTV}
\label{sec:ttv_predict}
\begin{figure}
 	\includegraphics[width=\columnwidth]{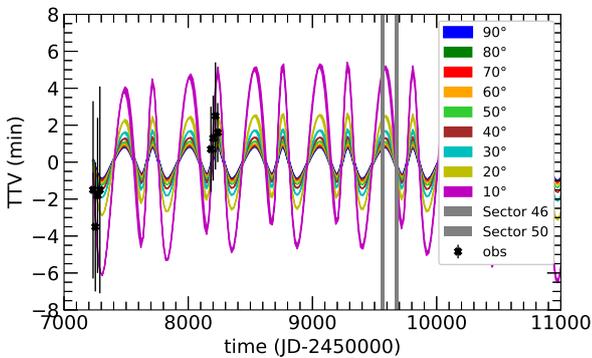}
     \caption{Predicted transit timing variations (TTVs) for \planetb, for a range of values of the orbital inclination of \planetc (coloured lines). The line thickness represents the uncertainties in the expected TTVs based on uncertainties in the planet masses and orbit parameters taken from Table~\ref{tab:mcmc}. Our measured mid-transit times (Table~\ref{tab:ttv} are shown as black crosses, and the time of future TESS observations of \target (Sectors 46 \& 50) are indicated with vertical shaded regions (see Section~\ref{sec:tess}).}
     \label{fig:ttv}
 \end{figure}

The presence of additional planets in a system containing a transiting planet is known to induce variations in the timings of the transits. Analysing these TTVs can lead to important insights into system architecture, such as inferring the presence of additional planets, and measuring planet masses (see e.g. \citealt{Agol_TTV} for a review). In order to compute the theoretically expected TTVs induced by the presence of \planetc, we carried out n-body simulations using \textsc{rebound} \citep{rebound_2012}, using the values reported in Table~\ref{tab:mcmc} for the planet and orbit parameters. Our simulations predict a TTV amplitude of around two to three minutes for most possible values of $i_\mathrm{c}$ (Fig.~\ref{fig:ttv}). The predicted TTVs are substantially larger for nearly face-on orbits ($i_\mathrm{c} = 10\degr$ is the smallest value included in Fig.~\ref{fig:ttv}), but
TTVs of even this magnitude cannot be ruled out using our measured transit times, because the uncertainties are too large (Table~\ref{tab:ttv}).

\subsubsection{Measured transit times}
\label{sec:ttv_obs}

We fitted for the times of each of the eight observed transits of planet `b'. This was done by fixing all parameters, except the transit epoch, to the best-fitting values listed in Table~\ref{tab:mcmc}. \tlcm was then used to fit only a single transit taken from the {\sc k2sc} long-cadence light curve (C6) or the short-cadence light curve described in Section~\ref{sec:C17} (C17). The resulting transit times and one-sigma uncertainties are listed in Table~\ref{tab:ttv}. We note that the uncertainties on the C6 transit times are somewhat larger than those reported in \paperone, and attribute this difference to the change of error estimation method within \tlcm (See Section~\ref{sec:results:planetb}).

We first compared the observed transit times to those predicted using the orbital ephemeris from Paper I, finding that all the C17 transits occur between 84 and 89 minutes earlier than predicted. However, the 1-sigma uncertainties on the times predicted by that ephemeris are around 80 minutes. Using the ephemeris reported in Table~\ref{tab:mcmc}, we find a maximum difference between predicted and observed transit times of 2.5 minutes. Given that the magnitude of the O-C differences are very similar to the timing precision of each transit, we conclude that there is no evidence for any deviation from a linear transit ephemeris. We also note that given the expected TTV amplitude (Section~\ref{sec:ttv_predict}) has a magnitude similar to the precision of our measurements, we do not expect to be able to measure the TTV signal.

\begin{table}
\caption{Fitted times of mid-transit for individual transits of \planetb, their uncertainties (in days and in minutes), and the deviations (O-C) from the ephemerides presented in \paperone and in Table~\ref{tab:mcmc}.}
\begin{tabular}{cccccc}\hline
$E$ &$T_{\rm c} - 2\ 450\ 000$ & $\sigma_{T_{\rm c}}$ & $\sigma_{T_{\rm c}}$ & \multicolumn{2}{c}{(O-C) / min} \\
&$\mathrm{BJD_{TDB}}$ & d & min & Paper I & Table~\ref{tab:mcmc} \\
\hline
0 & 7233.8264 & 0.0033 & 4.8 & 4.9    & -1.5 \\
1 & 7252.0730 & 0.0024 & 3.5 & 1.5    & -3.3 \\
2 & 7270.3218 & 0.0029 & 4.2 & 1.2    & -1.8 \\
3 & 7288.5698 & 0.0039 & 5.6 & -0.3   & -1.5 \\
52 & 8182.7125 & 0.0016 & 2.3 & -84.2  & 0.7 \\
53 & 8200.9607 & 0.0016 & 2.3 & -85.4  & 1.3 \\
54 & 8219.2094 & 0.002  & 2.9 & -85.9  & 2.5 \\
55 & 8237.4565 & 0.0011 & 1.6 & -88.5  & 1.6 \\
\hline
\\
\end{tabular}
\label{tab:ttv}
\end{table}

\subsection{Occultation}

We find no evidence for the occultation of planet `b' in the light curve of \target. By fitting for an occultation at the orbital phase expected from the $e_\mathrm{b}$ and $\omega_\mathrm{b}$ in Table~\ref{tab:mcmc} using \tlcm, we place a $3\sigma$ upper limit on the occultation depth of 130~ppm. That the occultation is not detected is unsurprising, given the large luminosity ($8\pm0.5$~\lsol) of the host star, and the moderate planetary equilibrium temperature.

\subsection{Additional periodicity in the RVs}

After subtracting the best-fitting Keplerian models for both planets, and the fitted instrumental offsets from our RVs, we searched the residuals for the presence of additional periodic signals, using the {\sc Astropy} \citep{astropy1,astropy2} implementation of the Lomb-Scargle periodogram. We found no further periodic signals with a false-alarm probability less than 10 per cent, and conclude that the RVs offer no evidence for the presence of a third planet, or for stellar rotation.

We also tried fitting for a radial acceleration term in our joint modelling with \tlcm, which returned $\dot\gamma=0.0144\pm0.0046$~m~s$^{-1}$~d$^{-1}$. The Bayesian Information Criterion (BIC) of the model with non-zero $\dot\gamma$ is lower by 7.8 which, along with the $3\,\sigma$ detection of $\dot \gamma$, could be interpreted as evidence in favour of the presence of an RV trend. However, we see no evidence of a power excess at low frequencies in the Fourier transform of the RV residuals. Such an excess is expected in the presence of a genuine radial acceleration, and a simple simulation using our RV timestamps and uncertainties confirms this. We further note that there may be a degeneracy between the radial acceleration term and the offsets between instruments \citep{knutson14}. The apparent presence of the trend is also strongly reliant on the most-recent HARPS observation. We conclude that the evidence for the presence of an RV trend is not compelling, and therefore choose to adopt the model with no radial acceleration term. We note that the two models are very similar to each other; no parameters vary between the models by more than $1\,\sigma$.

If the apparent acceleration term were real, it could indicate the presence of a third planet in orbit around \target. Following the same approach as we did for the much larger acceleration detected in \paperone, we find that $M_\mathrm{d} / a_\mathrm{d}^2 > 0.08$~\mjup~au$^{-2}$. If we assume that the orbit of the putative planet `d' is not highly eccentric, then the orbital period, $P_\mathrm{d}$ must be at least twice the baseline of our RV observations, $P_\mathrm{d} > 2294$~d which leads to a limit on the size of the orbit, $a_\mathrm{d} > 3.85$~au. A 2~\mjup planet orbiting at 5~au or a brown dwarf at several tens of au could induce an acceleration of this magnitude. Finally, we note that further radial velocity observations during \target's next observing season will enable the model with a radial accleration to be ruled out with greater certainty.

\subsection{Frequency analysis of the activity indicators}

We further assessed the planetary nature of the Doppler signal at 523~d by performing a frequency analysis of the bisector inverse slope (BIS) and FWHM of the cross-correlation function (CCF), as well as of Mount Wilson $S$-index ($S_\mathrm{MW}$). We used the generalised Lomb-Scargle periodogram \citep{Z&K}. We found no significant signal with a false-alarm probability \footnote{We estimated the false-alarm probability using the bootstrap method described in \cite{Murdoch1993}.} (FAP) lower than $\sim$8\,per cent. This acts as a sanity check that the RV signal with a period of 523~d is due to the Doppler reflex motion induced by an additional planet orbiting K2-99. We also used {\sc serval} \citep{Zechmeister2018} to measure additional activity indicators, namely, the chromatic index, differential line width, and H$\alpha$, sodium Na\,D1 \& Na\,D2 indexes. We found no significant periodic signal (FAP < 0.1\, per cent) in any of these indicators either, further corroborating our results.

\section{Possible transits of \planetc}
\label{sec:transits_c}
\subsection{The probability of transits}

Ignoring the presence of the inner planet, we can calculate the \textit{a priori} probability that the system is aligned such that the outer planet transits. This probability, calculated using Eqn. 9 of \cite{Winn_book}, but neglecting the planetary radius, such that a transit is defined as events where the star is occulted by at least half of the planetary disk, is 0.9 per cent. However, given the apparent propensity for at least some multi-planetary systems to have low mutual inclination angles (e.g. \citealt{Fabrycky14}), we might reasonably expect this probability to be higher.

\begin{figure}
 	\includegraphics[width=\columnwidth]{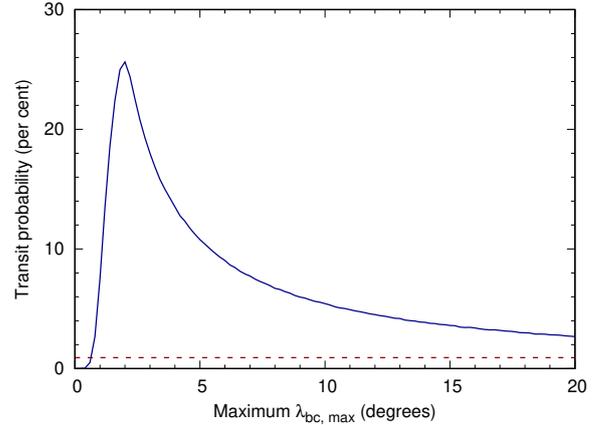}
     \caption{Probability that \planetc transits, as a function of the maximum mutual inclination angle between the orbital planes of planets `b' and `c', $\lambda_{\mathrm bc, max}$ (solid blue line). The \textit{a priori} transit probability of 0.92 per cent is shown as a dashed red line.}
     \label{fig:tr_prob}
 \end{figure}

The range of orbital inclination angles, $i_c$, for the outer planet that result in transits is $i_c \ge 89.478\pm0.007$ degrees. Following an approach similar to that of \cite{Beatty_Seager} for HAT-P-13\,b and \cite{Espinoza19} for GJ\,357\,d, we calculate the probability that \planetc transits for a range of mutual inclination angles (Fig.~\ref{fig:tr_prob}). We randomly draw values of the following parameters from normal distributions centred on the values listed in Tables~\ref{tab:stellar} and \ref{tab:mcmc}, and with standard deviations equal to the error bars listed in the same table: \rstar, \mstar, $i_\mathrm{b}$, $P_\mathrm{c}$, $e_\mathrm{c}$, $\omega_\mathrm{c}$. We then compute $i_\mathrm{c}$ by taking the drawn value of $i_\mathrm{b}$ and adding the mutual inclination angle, $\lambda_{\mathrm bc}$, which is drawn from a distribution which is uniform between $-\lambda_{\mathrm bc, max}$ and $\lambda_{\mathrm bc, max}$. This process is repeated a number of times for a single value of $\lambda_{\mathrm bc, max}$ to calculate a transit probability, before $\lambda_{\mathrm bc, max}$ is incremented and the whole process repeated.

As expected, the transit probability is significantly enhanced above the \textit{a priori} probability for small mutual inclinations, with a transit probability of just over 10 per cent for mutual inclinations less than $5\degr$, rising to more than 20 per cent for $\lambda_{\mathrm bc, max} = 2\degr$.

One of the results from the original \kep mission \citep{Kepler} is that tightly-packed multi-planet systems exhibit a flat geometry, with \cite{Fabrycky14} finding typical mutual inclination angles between orbital planes of just one or two degrees. However, the \target system is not representative of the tightly-packed systems of relatively small planets that dominate the \kep sample. Indeed, the so-called `\kep dichotomy', first observed by \cite{Lissauer11}, suggests that there is an excess of systems where only a single planet is observed to transit.

The \kep dichotomy has been interpreted as evidence for two populations of systems: flat multi-planet systems with low mutual inclinations, and a second population consisting of either lone planets, or multi-planet systems with high mutual inclinations. \cite{Albrecht13} and \cite{Morton+Winn14} found that the stellar obliquities of single transiting planet systems are systematically larger than for systems with multiple transiting planets, suggesting that single transiters represent a dynamically hotter population. The obliquity of \planetb is unknown, but if \target belongs to this latter population, like Kepler-108 which has two giant planets on 49-d and 190-d orbits and a mutual inclination angle of $24\substack{+11 \\ -8}\degr$ \citep{Mills+Fabrycky17} then it is probably rather unlikely that \planetc transits. 

More recently \cite{Zhu18} and \cite{Millholland21} have proposed models that explain the apparent \kep dichotomy with a continuous distribution of relatively small mutual inclinations. Here, the mutual inclination depends strongly on the intrinsic multiplicity of the system, such that systems that contain more planets are geometrically flatter, and there is no true dichotomy.

\subsection{The observability of transits}

Motivated by the previous section, in which we determine that there is a small, but significant chance that \planetc transits, we calculate the epoch of mid-transit for `c', assuming that its orbital inclination, $i_\mathrm{c}$, is equal to 90\degr. This epoch of mid-transit (see Table~\ref{tab:mcmc}), along the $P_\mathrm{c}$, allows us to calculate a list of potential future transits of the outer planet (Table~\ref{tab:transits_c}). The RA of \target results in an observing season that is centred on April, so the target will be very well observable during the next potential transit (in 2022 April).

Unfortunately, the uncertainty on this mid-transit time is relatively large, and so a photometric monitoring campaign of several weeks' duration would be required to cover the transit window. Given the mass of \planetc (Table~\ref{tab:mcmc}), we expect its radius and hence transit depth to be similar to those of the inner planet (around 0.2 per cent). The duration of a central ($i_\mathrm{c} = 90^{\circ}$, $b=0$) transit is calculated to be $32.1 \pm 0.5$ hours, with ingress and egress durations of around 1.3 hours, but these durations would be somewhat shorter for a transit with a moderate impact factor.

\begin{table}
\caption{Predicted times of mid-transit for \planetc, if it is transiting, from the ephemeris presented in Table~\ref{tab:mcmc}.}
\begin{tabular}{cccl}\hline
$E$ &$T_{\rm c} - 2\ 400\ 000$ & $\sigma_{T_{\rm c}}$ & Date \\
&$\mathrm{BJD_{TDB}}$ & d & \\
\hline
3 & 59670.7 & 4.6 & 2022 April 1 \\
4 & 60192.9 & 5.9 & 2023 September 5 \\
5 & 60715.1 & 7.3 & 2025 February 8 \\
6 & 61237.3 & 8.7 & 2026 July 15 \\
\hline
\\
\end{tabular}
\label{tab:transits_c}
\end{table}

\subsubsection{Prospects for observing the transit of \planetc from the ground}

The combination of depth and duration make this an extremely challenging transit to detect from the ground. Nevertheless, the ground-based Next Generation Transit Survey (NGTS; \citealt{NGTS}) has recently demonstrated its ability to detect extremely shallow ($< 0.1$ per cent) transits using several telescopes in combination \citep{NGTS_multi}. NGTS has also proven its ability to recover the transits of long-period objects by monitoring lengthy transit windows \citep{Gill_NGTS_EB,ngts11,Bryant21}. It may therefore be possible to detect a transit of \planetc with NGTS, although it would require committing several telescopes to observe for a period of some weeks.

\subsubsection{Prospects for observing the transit of \planetc from space}
\label{sec:tess}

The large transit window necessitates an infeasibly large time commitment for a targeted space-borne telescope, such as the CHaracterising ExOPlanet Satellite (CHEOPS; \citealt{CHEOPS}). The Transiting Exoplanet Survey Satellite (TESS; \citealt{TESS}) has not observed \target to date, however observations are expected\footnote{According to the Web TESS Viewing Tool; \url{https://heasarc.gsfc.nasa.gov/cgi-bin/tess/webtess/wtv.py}} during TESS' fourth year of operations (Sectors 46 and 50). No transit of \planetc is expected during S46 (2021 December), but the S50 observations are serendipitously timed to cover almost all of the 2022 April transit window. Current plans are for this sector to be observed from 2022 March 26 to 2022 April 22, meaning that all of the 9-d long one-sigma transit window, and much of the 18-d long two-sigma transit window will be observed. These planned TESS observations surely represent the best possibility of detecting the transit of \planetc.

Each TESS sector consists of two orbits of the satellite, between which there is a gap at perigee when the spacecraft is oriented for data downlink, and no observations are made. These gaps have a duration between 22 and 40 hours (based on Sectors 1-35). This pause in TESS observations around 2022 April 8/9, as well as a few days before the start of TESS observations would ideally be filled by CHEOPS observations, or by ground-based photometric observations from multiple longitudes, so that the transit is not missed should it occur during this time.

\subsection{The value of transits}

Detecting the transit of \planetc would be a valuable discovery, allowing measurement of its radius. Cold Jupiters are not subject to extreme insolation and tidal heating, allowing planetary evolution and interior models to be tested when planetary masses and radii are known. Relatively few such objects have been discovered so far; only 13 transiting planets with an orbital period longer than 500~d are currently listed in the NASA Exoplanet Archive\footnote{\url{https://exoplanetarchive.ipac.caltech.edu/}, accessed 2021 May 07.}. Of these 13, only eight have well-determined orbital periods, with the remainder having period uncertainties of greater than 20 per cent, or upper limits only. The eight transiting planets with well-determined periods greater than 500 d are all \kep targets, with no radial velocity measurements of the planetary mass. Indeed, only one transiting planet, the circumbinary Kepler-47c \citep{Kepler47}, is known with an orbital period greater than 300~d, and a well-determined mass, radius, and period. If a transit of \planetc is detected, it would be only the sixth exoplanet more massive than Saturn with measured mass, radius, and orbital period greater than 100~d, making it an extremely valuable object for further study.

\section{System architecture}
\label{sec:architecture}

\subsection{Orbital inclination of \planetc}

As we discussed in Section~\ref{sec:transits_c}, the inclination angle of the outer planet's orbit, $i_\mathrm{c}$, is unknown. We note that if the orbital inclination of the outer planet is less than about 40\degr, its mass would be above the deuterium burning limit of approximately 13~\mjup, placing it in the brown-dwarf mass regime. The evolutionary models of \cite{Petrovich&Tremaine} suggest that if \planetb is undergoing high-eccentricity migration, then \planetc is likely to have a large mutual inclination.

Astrometry in combination with RVs has the potential to fully solve the orbit of an exoplanet, determining the inclination angle, and hence the true planetary mass (e.g. \citealt{benedict02}). The final data release from {\it Gaia} is expected to enable this for a significant number of systems, as well as allowing the discovery of new exoplanets \citep{perryman14}. Even before this data release, for some systems the excess astrometric noise parameter, $\epsilon$, can enable constraints to be placed on the inclination angle \citep{kiefer19a,kiefer19b,kiefer21}. Unfortunately given the distance of \target, the size of the astrometric orbit of \planetc is too small to allow a meaningful constraint on $i_\mathrm{c}$, despite the small value of $\epsilon = 0.29$~mas in {\it Gaia} DR1 (F. Kiefer, private communication).

\subsection{A dynamically `hot' system?}

As we pointed out in \paperone, a measurement of the obliquity of \planetb ($\lambda_\mathrm{b}$) would be very interesting. This remains the case, and would also offer insight into whether or not the system is dynamically `hot', as discussed in Section~\ref{sec:transits_c}. Asteroseismology can offer a means to determine the inclination of the stellar rotation axis, and hence the obliquity of a planet for which the inclination angle is known. This technique \citep{Chaplin13} relies on a light curve with a signal-to-noise greater than that of our C17 light curve, and so we are not able to detect an asteroseismic rotation signature. The best prospect for measuring $\lambda_\mathrm{b}$ is therefore probably via the Rossiter-McLaughlin effect.

\subsection{Alignment of eccentricities}
\label{sec:apsidal}

\begin{figure}
 	\includegraphics[width=\columnwidth]{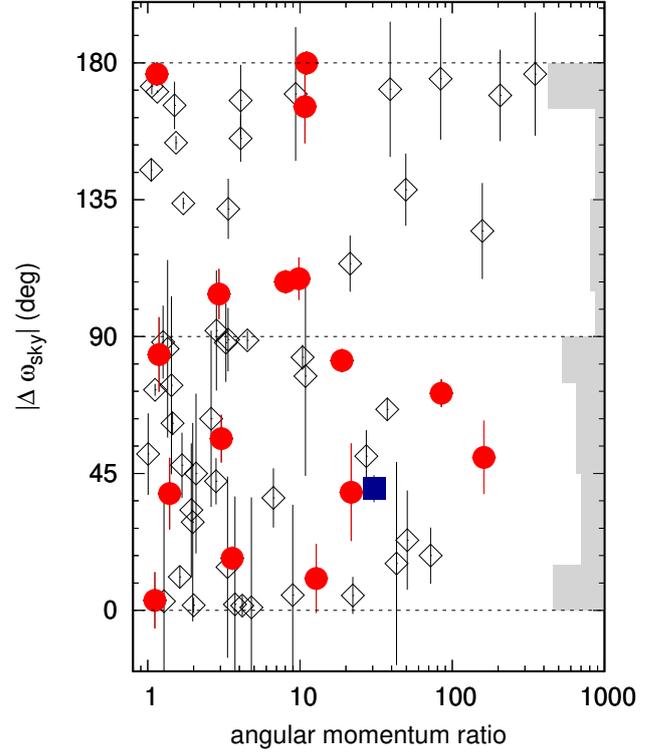}
     \caption{Sky-projected apsidal alignment as a function of orbital angular momentum ratio for a total of 67 pairs of planets in 45 systems, after \protect\cite{Dawson_Chiang}. The red circles represent warm Jupiters with a single outer companion orbiting beyond 1~AU. \target is shown by a blue square. The distribution of apsidal alignments for all points is shown as a grey histogram on the right hand side of the plot. Data is from the NASA Exoplanet Archive (see Section~\ref{sec:apsidal} for full details).}
     \label{fig:dawson}
 \end{figure}

Both planets in the \target system have similar orbital eccentricities ($e_\mathrm{b} = 0.22\pm0.02$ and $e_\mathrm{c} = 0.211\pm0.009$), although the apses of the two orbits are not aligned, $\left|\Delta \omega_\mathrm{sky} \right| = \left | \omega_\mathrm{b} - \omega_\mathrm{c} \right | = 40\degr \pm 4.5\degr$. \cite{Dawson_Chiang} studied warm Jupiters on eccentric orbits with giant companions also on an eccentric orbit. They found that in such systems, the orbits of the two planets tend to be apsidally misaligned, with an apparent clustering around $\left|\Delta \omega_\mathrm{sky} \right| = 90\degr$. Motivated by the apparent discrepancy of \target with this observation, we generated an updated version of Figure~1A of \cite{Dawson_Chiang} (Fig.~\ref{fig:dawson}). We used the same definitions and thresholds as \cite{Dawson_Chiang}, namely plotting pairs of planets where each planet has a $2\sigma$ detection of eccentricity, and $\sigma_\omega < 40\degr$. Using the NASA Exoplanet Archive, we calculated $\left|\Delta \omega_\mathrm{sky} \right|$ and the orbital angular momentum ratio (orbital angular momentum = $M_\mathrm{p} \sin i_\mathrm{p} \sqrt{a_\mathrm{p} (1-e_\mathrm{p}^2)}$).

Our sample is significantly larger than that available to \cite{Dawson_Chiang}; in Fig.~\ref{fig:dawson} we plot 67 pairs of planets, including 17 warm Jupiters with a single outer companion, compared to 40 pairs and eight warm Jupiters in \cite{Dawson_Chiang}. With this larger sample, we find no evidence for a clustering of apsidal alignment around 90\degr for the warm Jupiter systems. Fig.~\ref{fig:dawson} does however indicate that an alignment of orbital apses, such that $\left|\Delta \omega_\mathrm{sky} \right| \approx 0\degr$ and $\left|\Delta \omega_\mathrm{sky} \right| \approx 180\degr$ may be favoured for systems in general. Fig.~\ref{fig:dawson} also reveals that there are more systems with $\left|\Delta \omega_\mathrm{sky} \right| < 90\degr$ than with $\left|\Delta \omega_\mathrm{sky} \right| > 90\degr$. A two-sided binomial test (neglecting uncertainties in $\left|\Delta \omega_\mathrm{sky} \right|$) gives $p=0.036$, indicating that this imbalance is significant at the five per cent level. However, when points whose 1\,$\sigma$ error bars straddle the $90\degr$-line are neglected, we find $p=0.23$ which offers no evidence in favour of an imbalance in the distribution of $\left|\Delta \omega_\mathrm{sky} \right|$.

\section{Conclusions}
\label{sec:conclusions}

We have used newly-obtained RVs to measure the mass, and determine the orbit of \planetc. We refined our knowledge of \target the star via Gaia parallax measurements, and an asteroseismic analysis of short-cadence photometry from \ktwo's Campaign 17. We also improved our knowledge of the inner planet, \planetb, finding it to be non-inflated. Upcoming TESS observations of \target in 2022 April offer the tantalising possibility of detecting the transit of \planetc, if the orbital planes of the two planets have a small mutual inclination angle. Alternatively, if the mutual inclination angle is large, \planetc may have played an important role in the inward migration of \planetb, making the \target system an important laboratory for our understanding of planetary migration. The full three-dimensional geometry of the system may be revealed in the future through a measurement of the spin-orbit alignment of \planetb and a measurement of the inclination of \planetc via detection of its transit, or from astrometry.

\section*{Acknowledgements}
This work is done under the framework of the KESPRINT collaboration (\url{http://www.kesprint.science}). KESPRINT is an international consortium devoted to the characterization and research of exoplanets discovered with space-based missions.

Based on observations made with ESO Telescopes at the La Silla Observatory (Chile) under programmes 097.C-0948(A), 099.C-0491(B), 099.C-0491(A), 0100.C-0808(A), 0101.C-0829(A), 60.A-9700(G), and 1102.C-0923(A), and the TNG telescope at Roque de Los Muchachos Observatory (Spain) under programmes A33TAC\_15, A34TAC\_10, OPT17A\_64, A35TAC\_26, OPT17B\_59, CAT17B\_99, CAT18A\_130, OPT18A\_44, A37TAC\_37, OPT18B\_52, and A38TAC\_26.

This paper includes data collected by the \kep mission. Funding for the \kep mission is provided by the NASA Science Mission directorate. Some of the data presented in this paper were obtained from the Mikulski Archive for Space Telescopes (MAST). STScI is operated by the Association of Universities for Research in Astronomy, Inc., under NASA contract NAS5-26555. Support for MAST for non-HST data is provided by the NASA Office of Space Science via grant NNX09AF08G and by other grants and contracts.

This research has made use of the Exoplanet Follow-up Observation Program website, and the NASA Exoplanet Archive, which are operated by the California Institute of Technology, under contract with the National Aeronautics and Space Administration under the Exoplanet Exploration Program.

The research leading to these results has received funding from the European Union Seventh Framework Programme (FP7/2013-2016) under grant agreement No. 312430 (OPTICON).

This research has made use of NASA's Astrophysics Data System, the SIMBAD data base, operated at CDS, Strasbourg, France, and the Exoplanets Encyclopaedia at exoplanet.eu. We also used Astropy, a community-developed core Python package for Astronomy \citep{astropy1,astropy2}.

This work has made use of data from the European Space Agency (ESA) mission {\it Gaia} (\url{https://www.cosmos.esa.int/gaia}), processed by the {\it Gaia} Data Processing and Analysis Consortium (DPAC, \url{https://www.cosmos.esa.int/web/gaia/dpac/consortium}). Funding for the DPAC
has been provided by national institutions, in particular the institutions participating in the {\it Gaia} Multilateral Agreement.


K.W.F.L., Sz.Cs., and A.P.H. were supported by Deutsche Forschungsgemeinschaft grants HA3279/12-1 and RA714/14-1 within the DFG Schwerpunkt SPP 1992, Exploring the Diversity of Extrasolar Planets. Sz.Cs. is also supported by Deutsche Forschungsgemeinschaft Research Unit 2440: ’Matter Under Planetary Interior Conditions: High Pressure Planetary and Plasma Physics’. H.J.D. acknowledges support from the Spanish Research Agency of the Ministry of Science and Innovation (AEI-MICINN) under grant PID2019-107061GB-C66, DOI: 10.13039/501100011033. J.K. gratefully acknowledge the support of the Swedish National Space Agency (SNSA; DNR 2020-00104). S.M. acknowledges support by the Spanish Ministry of Science and Innovation with the Ramon y Cajal fellowship number RYC-2015-17697 and the grant number PID2019-107187GB-I00. S.N.B. and R.A.G. acknowledge the support of the PLATO CNES grant. L.M.S. and D.G. gratefully acknowledge financial support from the CRT foundation under Grant No. 2018.2323 ``Gaseous or rocky? Unveiling the nature of small worlds".
We thank Erik Petigura for his contribution to the collection and analysis of the Keck/HIRES data. Finally, we thank the referee for their careful reading of the manuscript and constructive suggestions, which resulted in improvements to this paper.

\section*{Data availability}

The data underlying this article are available in the article and at the following public archives. The HARPS spectra can be found at the ESO archive\footnote{\url{http://archive.eso.org/}}; the HARPS-N spectra at the TNG archive\footnote{\url{http://archives.ia2.inaf.it/tng/}}; the HIRES spectra at EXOFOP-K2\footnote{\url{https://exofop.ipac.caltech.edu/k2/}}; and the \ktwo data at MAST\footnote{\url{https://archive.stsci.edu/}}.

\bibliographystyle{mnras}
\bibliography{refs2} 


\appendix

\section{Asteroseismic frequencies}

\begin{table*}
\caption{Asteroseismic frequencies of the modes fitted in the PSD with {\sc apollinaire}.}

\begin{tabular}{ccc}\hline
Radial order, $n$ & Spherical degree, $\ell$ & Frequency ($\mu$Hz) \\
\hline
13 & 2 & 611.89 $\pm$ 1.64\\
14 & 0 & 616.15 $\pm$ 1.12\\
14 & 1 & 631.97 $\pm$ 0.77\\
15 & 0 & 655.15 $\pm$ 0.78\\
15 & 1 & 672.34 $\pm$ 0.46\\
15 & 2 & 690.64 $\pm$ 2.05\\
16 & 0 & 694.07 $\pm$ 1.02\\
16 & 1 & 712.34 $\pm$ 0.88\\
16 & 2 & 732.17 $\pm$ 2.41\\
17 & 0 & 735.71 $\pm$ 0.48\\
17 & 1 & 753.18 $\pm$ 0.39\\
17 & 2 & 772.20 $\pm$ 0.99\\
18 & 0 & 777.09 $\pm$ 0.63\\
18 & 1 & 795.28 $\pm$ 0.39\\
18 & 2 & 813.88 $\pm$ 3.27\\
19 & 0 & 817.39 $\pm$ 0.54\\
19 & 1 & 835.30 $\pm$ 1.62\\
19 & 2 & 853.06 $\pm$ 1.83\\
20 & 0 & 856.21 $\pm$ 1.03\\
20 & 1 & 974.97 $\pm$ 1.73\\
\hline
\\
\end{tabular}
\label{tab:astero}
\end{table*}

\section{Radial velocity data}

\begin{table*}
\caption{Radial velocity (RV) measurements of \target, from the HARPS and HARP-N instruments. BIS is the bisector span, FWHM is the full width at half maximum of the cross-correlation function, $S_\mathrm{MW}$ is the Mount Wilson $S$-index, S/N is the signal-to-noise, and $T_\mathrm{exp}$ is the exposure time. Measurements marked with $\dagger$ were previously presented in \paperone.}
\begin{tabular}{lrrrrrrrrrl}\hline
$\mathrm{BJD_{TDB}}$ & RV & $\sigma_{\mathrm{RV}}$ & BIS & $\sigma_\mathrm{BIS}$  & FWHM & $S_\mathrm{MW}$ & $\sigma_{S_\mathrm{MW}}$ & S/N & $T_\mathrm{exp}$ & Instrument \\
 $-2450000$ & \kms & \kms  &  \ms & \ms & \kms & & & @550~nm & s & \\
\hline
7492.520141  &  -2.5811  &  0.0084  &   -7.355 & 11.839 &  14.568 & 0.1180 &  0.0068 &  36.6 &  1500.0  &  HARPS-N $^\dagger$ \\    
7502.643805  &  -2.4930  &  0.0074  &  -31.332 & 10.427 &  14.481 & 0.1210 &  0.0059 &  39.2 &  1500.0  &  HARPS-N $^\dagger$ \\    
7512.508450  &  -2.6160  &  0.0048  &    6.506 &  6.733 &  14.598 & 0.1327 &  0.0031 &  60.7 &  1560.0  &  HARPS-N $^\dagger$ \\    
7532.518735  &  -2.6452  &  0.0059  &  -34.171 &  8.348 &  14.600 & 0.1221 &  0.0041 &  50.5 &  2400.0  &  HARPS-N $^\dagger$ \\    
7539.461243  &  -2.5584  &  0.0045  &  -18.941 &  6.409 &  14.566 & 0.1306 &  0.0030 &  63.5 &  2400.0  &  HARPS-N $^\dagger$ \\    
7844.622488  &  -2.8544  &  0.0081  &  -38.028 & 11.501 &  14.611 & 0.1390 &  0.0061 &  38.7 &  2400.0  &  HARPS-N            \\    
7852.569130  &  -2.7689  &  0.0050  &   11.587 &  7.074 &  14.616 & 0.1380 &  0.0034 &  57.8 &  2700.0  &  HARPS-N            \\    
7900.496311  &  -2.7627  &  0.0033  &   -2.269 &  4.737 &  14.582 & 0.1282 &  0.0024 &  84.1 &  3360.0  &  HARPS-N            \\    
7924.411710  &  -2.6385  &  0.0031  &  -12.979 &  4.402 &  14.587 & 0.1307 &  0.0027 &  86.8 &  4000.0  &  HARPS-N            \\    
7938.407862  &  -2.6539  &  0.0033  &  -18.759 &  4.650 &  14.577 & 0.1330 &  0.0028 &  82.0 &  3420.0  &  HARPS-N            \\    
7939.417617  &  -2.6298  &  0.0043  &  -41.259 &  6.096 &  14.573 & 0.1248 &  0.0029 &  65.6 &  2700.0  &  HARPS-N            \\    
7958.399288  &  -2.5799  &  0.0055  &   11.055 &  7.749 &  14.581 & 0.1257 &  0.0036 &  52.5 &  3420.0  &  HARPS-N            \\    
7965.393368  &  -2.6214  &  0.0032  &    4.086 &  4.525 &  14.588 & 0.1258 &  0.0026 &  87.2 &  3420.0  &  HARPS-N            \\    
8114.772167  &  -2.7847  &  0.0057  &  -15.264 &  8.059 &  14.524 & 0.1341 &  0.0048 &  49.8 &  2100.0  &  HARPS-N            \\    
8140.737815  &  -2.7458  &  0.0123  &   -8.882 & 17.387 &  14.697 & 0.1720 &  0.0224 &  30.0 &  1500.0  &  HARPS-N            \\    
8140.763315  &  -2.7409  &  0.0101  &  -24.925 & 14.229 &  14.641 & 0.1409 &  0.0164 &  34.2 &  1500.0  &  HARPS-N            \\    
8168.664240  &  -2.8620  &  0.0090  &   12.135 & 12.685 &  14.577 & 0.1284 &  0.0074 &  34.9 &  2400.0  &  HARPS-N            \\    
8169.658260  &  -2.8811  &  0.0054  &  -10.557 &  7.704 &  14.576 & 0.1344 &  0.0038 &  54.1 &  3000.0  &  HARPS-N            \\    
8202.660478  &  -2.8848  &  0.0086  &  -35.568 & 12.186 &  14.588 & 0.1444 &  0.0062 &  36.7 &  3600.0  &  HARPS-N            \\    
8220.672326  &  -2.8936  &  0.0090  &   26.086 & 12.700 &  14.550 & 0.1233 &  0.0059 &  36.7 &  2100.0  &  HARPS-N            \\    
8227.524773  &  -2.9213  &  0.0050  &   -4.321 &  7.134 &  14.577 & 0.1335 &  0.0034 &  57.1 &  1800.0  &  HARPS-N            \\    
8286.464040  &  -2.8669  &  0.0064  &  -26.870 &  9.027 &  14.567 & 0.1218 &  0.0043 &  46.4 &  1800.0  &  HARPS-N            \\    
8289.464071  &  -2.8252  &  0.0041  &   -6.143 &  5.747 &  14.606 & 0.1294 &  0.0029 &  69.9 &  2100.0  &  HARPS-N            \\    
8289.490365  &  -2.8317  &  0.0043  &   -6.980 &  6.041 &  14.618 & 0.1265 &  0.0032 &  67.1 &  2100.0  &  HARPS-N            \\    
8313.393171  &  -2.8989  &  0.0050  &   -9.429 &  7.136 &  14.577 & 0.1182 &  0.0038 &  55.5 &  1800.0  &  HARPS-N            \\    
8313.415102  &  -2.9060  &  0.0058  &    0.911 &  8.246 &  14.596 & 0.1382 &  0.0040 &  49.4 &  1800.0  &  HARPS-N            \\  
8314.393515  &  -2.9108  &  0.0067  &    1.094 &  9.466 &  14.551 & 0.1283 &  0.0052 &  42.0 &  1800.0  &  HARPS-N\\
8493.789924  &  -2.6062  &  0.0059  &  -20.095 &  8.348 &  14.612 & 0.1372 &  0.0037 &  50.9 &  1500.0  &  HARPS-N\\
8552.631863  &  -2.5933  &  0.0165  &   42.798 & 23.289 &  14.669 & 0.1365 &  0.0144 &  21.6 &  1800.0  &  HARPS-N\\
7511.732775  &  -2.6017  &  0.0040  &  -40.191 &  5.615 &  14.527 & 0.1279 &  0.0038 &  70.0 &  1800.0  &  HARPS $^\dagger$\\
7512.635485  &  -2.6224  &  0.0037  &   -0.119 &  5.170 &  14.536 & 0.1366 &  0.0030 &  73.5 &  3600.0  &  HARPS $^\dagger$\\
7515.727289  &  -2.6171  &  0.0121  &   53.750 & 17.082 &  14.470 & 0.1480 &  0.0108 &  25.0 &  2271.2  &  HARPS $^\dagger$\\
7516.570134  &  -2.5982  &  0.0055  &  -53.411 &  7.719 &  14.521 & 0.1380 &  0.0030 &  48.6 &  2400.0  &  HARPS $^\dagger$\\
7559.602346  &  -2.6204  &  0.0050  &  -22.645 &  7.016 &  14.470 & 0.1357 &  0.0035 &  55.9 &  1800.0  &  HARPS $^\dagger$\\
7561.582108  &  -2.6492  &  0.0054  &  -14.259 &  7.670 &  14.484 & 0.1316 &  0.0034 &  50.6 &  1800.0  &  HARPS $^\dagger$\\
7589.496507  &  -2.7792  &  0.0059  &  -49.994 &  8.288 &  14.477 & 0.1448 &  0.0037 &  47.0 &  1800.0  &  HARPS $^\dagger$\\
7610.468853  &  -2.7700  &  0.0048  &  -25.429 &  6.822 &  14.483 & 0.1388 &  0.0033 &  57.2 &  1800.0  &  HARPS $^\dagger$\\
7984.477074  &  -2.6151  &  0.0063  &   -3.222 &  8.953 &  14.528 & 0.1279 &  0.0050 &  48.0 &  2700.0  &  HARPS\\
7987.480326  &  -2.6165  &  0.0060  &  -59.270 &  8.491 &  14.533 & 0.1357 &  0.0054 &  50.8 &  2100.0  &  HARPS\\
8143.843055  &  -2.7671  &  0.0067  &   -4.569 &  9.544 &  14.529 & 0.1353 &  0.0050 &  41.8 &  1800.0  &  HARPS\\
8171.862785  &  -2.9108  &  0.0056  &  -18.794 &  7.907 &  14.884 & 0.0466 &  0.1248 &  66.7 &  2400.0 &  HARPS\\
8172.875518  &  -2.8964  &  0.0053  &   -8.069 &  7.447 &  14.539 & 0.1245 &  0.0043 &  51.7 &  1800.0  &  HARPS\\
8173.893321  &  -2.8862  &  0.0043  &  -11.298 &  6.041 &  14.556 & 0.1356 &  0.0039 &  64.9 &  1800.0  &  HARPS\\
8191.819419  &  -2.9021  &  0.0048  &  -35.924 &  6.833 &  14.559 & 0.1325 &  0.0031 &  53.6 &  1800.0  &  HARPS\\
8192.846284  &  -2.8925  &  0.0043  &  -13.804 &  6.125 &  14.565 & 0.1371 &  0.0028 &  60.2 &  1800.0  &  HARPS\\
8220.726900  &  -2.9104  &  0.0064  &  -32.585 &  9.034 &  14.561 & 0.1186 &  0.0044 &  41.7 &  1800.0  &  HARPS\\
8249.704832  &  -2.8779  &  0.0064  &  -31.228 &  9.106 &  14.554 & 0.1250 &  0.0044 &  42.9 &  1800.0  &  HARPS\\
8250.716983  &  -2.8493  &  0.0045  &   -8.622 &  6.347 &  14.528 & 0.1326 &  0.0031 &  58.7 &  2400.0  &  HARPS\\
8251.704250  &  -2.8325  &  0.0050  &  -22.395 &  7.002 &  14.555 & 0.1401 &  0.0033 &  53.8 &  2100.0  &  HARPS\\
8324.532222  &  -2.8181  &  0.0069  &    1.516 &  9.715 &  14.545 & 0.1029 &  0.0041 &  40.4 &  2400.0  &  HARPS\\
8325.547112  &  -2.8132  &  0.0052  &  -17.536 &  7.380 &  14.535 & 0.1448 &  0.0033 &  52.4 &  2400.0  &  HARPS\\
8328.513677  &  -2.8534  &  0.0060  &  -16.618 &  8.446 &  14.522 & 0.1557 &  0.0051 &  46.2 &  2100.0  &  HARPS\\
8359.473765  &  -2.8121  &  0.0055  & -18.488 & 7.841 & 14.485 & 0.1387 & 0.0040 & 48.8 & 2400.0 &  HARPS\\
8626.733290  &  -2.7846  &  0.0085  & -3.922 & 12.058 & 14.639 & 0.1137 & 0.0070 & 37.7 & 1800.0 &  HARPS\\
\hline
\\
\end{tabular}
\label{tab:rvs1}
\end{table*}

\begin{table*}
\caption{Radial velocity (RV) measurements of \target, from the FIES, Tull, and HIRES instruments. $S_\mathrm{MW}$ is the Mount Wilson $S$-index, and $T_\mathrm{exp}$ is the exposure time. Measurements marked with $\dagger$ were previously presented in \paperone.}
\begin{tabular}{lrrrrrrl}\hline
$\mathrm{BJD_{TDB}}$ & RV & $\sigma_{\mathrm{RV}}$ & $S_\mathrm{MW}$ & $\sigma_{S_\mathrm{MW}}$ &  Texp & Instrument \\
 $-2450000$ & \kms & \kms  &  &   & s & \\
\hline
7479.624340  &  -2.6970  &  0.0134  &        &         &  3600.0  &  FIES  $^\dagger$   \\    
7503.531525  &  -2.6011  &  0.0137  &        &         &  3600.0  &  FIES  $^\dagger$   \\    
7523.478018  &  -2.6303  &  0.0188  &        &         &  3600.0  &  FIES  $^\dagger$   \\    
7565.410818  &  -2.8055  &  0.0157  &        &         &  3600.0  &  FIES  $^\dagger$   \\    
7566.413167  &  -2.7982  &  0.0138  &        &         &  3600.0  &  FIES  $^\dagger$   \\    
7567.416731  &  -2.8495  &  0.0144  &        &         &  3600.0  &  FIES  $^\dagger$   \\    
7568.417452  &  -2.8338  &  0.0182  &        &         &  4500.0  &  FIES  $^\dagger$   \\    
7570.405863  &  -2.8192  &  0.0162  &        &         &  3600.0  &  FIES  $^\dagger$   \\    
7572.408029  &  -2.8093  &  0.0157  &        &         &  2700.0  &  FIES  $^\dagger$   \\    
7575.409114  &  -2.7400  &  0.0182  &        &         &  2700.0  &  FIES  $^\dagger$   \\    
7576.403828  &  -2.7255  &  0.0152  &        &         &  2700.0  &  FIES  $^\dagger$   \\    
7577.404365  &  -2.7567  &  0.0197  &        &         &  3000.0  &  FIES  $^\dagger$   \\    
7578.405228  &  -2.7605  &  0.0161  &        &         &  3000.0  &  FIES  $^\dagger$   \\    
7579.402440  &  -2.8069  &  0.0217  &        &         &  3000.0  &  FIES  $^\dagger$   \\    
7493.757674  &   0.5015  &  0.0194  &        &         &          &  Tull $^\dagger$    \\    
7494.804635  &   0.4674  &  0.0158  &        &         &          &  Tull $^\dagger$    \\    
7524.768623  &   0.4463  &  0.0153  &        &         &          &  Tull $^\dagger$    \\    
7542.699191  &   0.4771  &  0.0077  &        &         &          &  Tull $^\dagger$    \\    
7543.736409  &   0.4330  &  0.0105  &        &         &          &  Tull $^\dagger$    \\    
7545.696704  &   0.4162  &  0.0206  &        &         &          &  Tull $^\dagger$    \\    
7561.891343  &   0.1038  &  0.0037  & 0.1194 &  0.001  &  459.5   &  HIRES\\
7579.754488  &   0.0721  &  0.0032  & 0.1215 &  0.001  &  321.9   &  HIRES\\
7579.776604  &   0.0554  &  0.0033  & 0.1212 &  0.001  &  370.3   &  HIRES\\
7579.792089  &   0.0740  &  0.0031  & 0.1184 &  0.001  &  351.9   &  HIRES\\
7586.786379  &  -0.0175  &  0.0034  & 0.1187 &  0.001  &  355.9   &  HIRES\\
7586.839997  &  -0.0269  &  0.0034  & 0.1199 &  0.001  &  397.3   &  HIRES\\
7587.773540  &  -0.0337  &  0.0033  & 0.1187 &  0.001  &  383.3   &  HIRES\\
7587.853324  &  -0.0245  &  0.0036  & 0.118  &  0.001  &  448.0   &  HIRES\\
7595.802575  &   0.0492  &  0.0037  & 0.1168 &  0.001  &  629.1   &  HIRES\\
7598.812828  &  -0.0037  &  0.0038  & 0.117  &  0.001  &  624.7   &  HIRES\\
7599.757760  &  -0.0115  &  0.0034  & 0.1214 &  0.001  &  386.6   &  HIRES\\
7600.786110  &  -0.0182  &  0.0034  & 0.1185 &  0.001  &  477.8   &  HIRES\\
7612.756877  &   0.0392  &  0.0036  & 0.1183 &  0.001  &  423.3   &  HIRES\\
7615.757684  &  -0.0007  &  0.0038  & 0.1221 &  0.001  &  462.0   &  HIRES\\
7616.757179  &  -0.0168  &  0.0036  & 0.1203 &  0.001  &  436.9   &  HIRES\\
7617.752589  &  -0.0517  &  0.0036  & 0.1197 &  0.001  &  425.6   &  HIRES\\
7618.747133  &  -0.0444  &  0.0037  & 0.1215 &  0.001  &  381.7   &  HIRES\\
7620.753303  &  -0.0676  &  0.0041  & 0.119  &  0.001  &  547.8   &  HIRES\\
7621.751529  &  -0.0747  &  0.0039  & 0.114  &  0.001  &  698.1   &  HIRES\\
\hline
\\
\end{tabular}
\label{tab:rvs2}
\end{table*}

\bsp	
\label{lastpage}
\end{document}